\documentclass[twocolumn,showpacs,aps,pre]{revtex4-1} 
\usepackage{amsmath}
\usepackage{graphicx}
\usepackage{color}
\usepackage{xcolor}
\usepackage{stmaryrd}
\usepackage{amssymb}
\usepackage{wasysym}
\usepackage[utf8]{inputenc}

\begin{document}

\title{Mach-like capillary-gravity wakes}

\author{Fr\'{e}d\'{e}ric Moisy}
\author{Marc Rabaud}
\affiliation{Universit\'e Paris-Sud, CNRS, Laboratoire FAST, B\^atiment 502, 91405 Orsay, France}

\date{\today}


\begin{abstract}

We determine experimentally the angle $\alpha$ of maximum wave amplitude in the far-field wake behind a vertical surface-piercing cylinder translated at constant velocity $U$ for Bond numbers $\mathrm{Bo}_D = D / \lambda_c$ ranging between 0.1 and 4.2, where $D$ is the cylinder diameter and $\lambda_c$ the capillary length. In all cases the wake angle is found to follow a Mach-like law at large velocity, $\alpha \sim U^{-1}$, but with different prefactors depending on the value of $\mathrm{Bo}_D$. For small $\mathrm{Bo}_D$ (large capillary effects), the wake angle approximately follows the law $\alpha \simeq c_{\rm g,min} / U$, where $c_{\rm g,min}$ is the minimum group velocity of capillary-gravity waves. For larger $\mathrm{Bo}_D$ (weak capillary effects), we recover a law $\alpha \sim \sqrt{gD}/U$ similar to that found for ship wakes at large velocity [Rabaud and Moisy, Phys. Rev. Lett. {\bf 110},  214503 (2013)].   Using the general property of dispersive waves that the characteristic wavelength of the wavepacket emitted by a disturbance is of order of the disturbance size, we propose a simple model that describes the transition between these two Mach-like regimes as the Bond number is varied. We show that the new capillary law $\alpha \simeq c_{\rm g,min} / U$ originates from the presence of a capillary cusp angle (distinct from the usual gravity cusp angle), along which the energy radiated by the disturbance accumulates for Bond numbers of order of unity. This model, complemented by numerical simulations of the surface elevation induced by a moving Gaussian pressure disturbance, is in qualitative agreement with experimental measurements.

\end{abstract}

\maketitle

\section{Introduction}

A ship moving on calm water generates gravity waves presenting a characteristic V-shaped pattern. Lord Kelvin~\cite{Kelvin1887} in 1887 was the first to describe the structure of this pattern,  using a stationary phase argument~\cite{Havelock1908,Lamb,Wehausen,Whitham,Lighthill,Darrigol}. The key result of Kelvin's analysis is that the energy radiated by the disturbance remains confined in a wedge of half-angle given by $\sin^{-1}(1/3) \simeq 19.47^\mathrm{o}$, independent of its size and velocity. Interestingly, although the geometry of the crest lines is universal,  different regions of the pattern may be visible or hidden depending on which wave numbers are effectively radiated by the disturbance or how excited waves produce constructive or destructive interferences~\cite{Wehausen,Lighthill,Raven2010,Rousseaux2013}. The full Kelvin pattern is visible only if all wavelengths are equally radiated by the disturbance, but in general only a finite range of wavelengths is visible, which affects the overall shape of the pattern.

To account for the large variety of wakes observed behind ships of different size and velocities, the details of the ship geometry and the nature of the flow around it must be considered. However, focusing on the far-field angle of maximum wave amplitude produced by a disturbance characterized by a single length scale $L$, the wake can be described solely by the Froude number $\mathrm{Fr} = U / \sqrt{gL}$, with $U$ the velocity and $g$ the gravitational acceleration, provided that the capillary effects can be neglected~\cite{Wehausen,Lighthill}. The dependence of the angle with $\mathrm{Fr}$, and the physical origin of this dependence, have recently received much attention~\cite{Rabaud2013,Darmon2014,Ellingsen2014,Noblesse2014,Moisy2014,Rabaud2014,Benzaquen2014,Pethi2014}. Note that considering a finite water depth $H$ introduces another Froude number, $Fr_H = U / \sqrt{gH}$, which has also a strong influence on the geometry of the wake pattern, but which we shall not consider in this paper.

In order to describe the far-field wake angle, we used in Refs.~\cite{Rabaud2013,Moisy2014} the key property of dispersive waves that the waves of maximum amplitude generated by a disturbance of size $L$ are of wavelength of order of $L$. This is the main result of the Cauchy-Poisson initial value problem \cite{Havelock1908,Lamb,Wehausen,Whitham,Lighthill}, which describes the evolution of the free surface elevation originating from an initial disturbance: The wave packet emitted by a disturbance of size $L$ propagates at the group velocity $c_g (k_f) = \frac{1}{2}\sqrt{g/k_f}$, where the local wave number $k_f$ at the maximum of the wave packet is of order of $L^{-1}$. Wave lengths much larger or much smaller than $L$ are found far from the maximum of the wave packet, and are therefore of weak amplitude. As a consequence, among the range of wavelengths necessary to build the full Kelvin pattern (between 0 and $\lambda_g = 2 \pi U^2/g$), only those of order of $L$ have significant amplitude, yielding an angle of maximum wave amplitude smaller than the Kelvin angle when $L \ll \lambda_g$ (i.e. when $\mathrm{Fr} \gg 1/\sqrt{2\pi} \simeq 0.4$). In this case the far-field wake angle is simply obtained by considering the superposition of wave packets emitted along the disturbance trajectory and propagating at constant group velocity $c_g (k_f)$, which plays the role of an effective sound velocity as in a non-dispersive medium. The wake angle is therefore given by a Mach-like law $\alpha \simeq \sin^{-1} (c_g (k_f)/U)$~\cite{cerenkov}, yielding at large Froude number
\begin{equation}
\alpha \simeq \frac{a}{\mathrm{Fr}},
\label{eq:am}
\end{equation}
with $a \simeq O(1)$. This scaling has been confirmed analytically by Darmon {\it et al.} \cite{Darmon2014} for an axisymmetric Gaussian pressure disturbance.
Equation~(\ref{eq:am}) does not apply at moderate Froude number, when $L$ is of order or larger than $\lambda_g$, for which most of the energy radiated by the disturbance concentrates along the cusp lines at the Kelvin angle. The value of $a$ in Eq.~(\ref{eq:am}) depends on the shape of the disturbance, which sets the relation between its characteristic size $L$ and the dominant wavenumber $k_f$ emitted in the wave packet. The simple choice $k_f = 2\pi/L$ proposed in Ref.~\cite{Rabaud2013} yields $a=1/(2 \sqrt{2\pi}) \simeq 0.20$, which turns out to provide a reasonable agreement with the wake angles measured from airborne images of ship wakes.

The case of non-axisymmetric pressure disturbance, recently discussed in Moisy and Rabaud~\cite{Moisy2014} and Benzaquen {\it et al.}~\cite{Benzaquen2014}, suggests that Eq.~(\ref{eq:am}) remains asymptotically valid provided that the Froude number is based on the {\it width} of the disturbance. Elongated pressure disturbances actually show a transition between an intermediate scaling $\alpha \simeq \mathrm{Fr}^{-2}$ and the asymptotic scaling $\alpha \simeq \mathrm{Fr}^{-1}$~\cite{Moisy2014}. This intermediate scaling $\alpha \simeq \mathrm{Fr}^{-2}$ has been first proposed by Noblesse {\it et al.}~\cite{Noblesse2014} by considering the interference between the Kelvin wakes emitted by two point sources separated by a distance $L$, aiming to model the bow and stern waves of a ship. In the range of Froude numbers for which ships are usually designed ($\mathrm{Fr} < 2$), both laws $\alpha \simeq \mathrm{Fr}^{-1}$ and $\mathrm{Fr}^{-2}$ are actually compatible with the available data. Larger Froude numbers, up to 5 or 10, may be encountered for offshore powerboats (or ``go-fast boats''), although no wake angle measurements are available in this regime to our knowledge.

Very large Froude numbers up to 10 are more commonly encountered for small objects, such as submarine periscopes, water skis, or hydrofoils. In these cases, the object size ($10-$ to $20-$cm diameter for a periscope and 3~cm~$\times$~30~cm cross-section for a small sailboat hydrofoil) is comparable to the capillary length, $\lambda_c = 2 \pi (\gamma / \rho g)^{1/2} \simeq 1.5$~cm, suggesting significant influence of the capillary effects ($\rho$ is the fluid density and $\gamma$ the surface tension).
The geometry of the capillary-gravity crest lines has been described in detail in Refs.~\cite{Lamb,Crapper1964,Binnie1965,Yih2,Doyle2013}. To account for the finite size effects of the disturbance, two non-dimensional numbers must be introduced: In addition to the Froude number $\mathrm{Fr}$, capillary-gravity wakes are also characterized by a Bond number,  $\mathrm{Bo} = L / \lambda_c$, or, equivalently, by the velocity ratio ${\cal U} = U / c_{\rm min}$, where $c_{\rm min} = (4 g \gamma / \rho)^{1/4} \simeq 22$~cm~s$^{-1}$ is the minimum phase velocity (the three non-dimensional numbers are related by ${\cal U} = \mathrm{Fr} \sqrt{\pi \mathrm{Bo}}$). Capillary-gravity wakes have been mostly investigated in connection with the wave resistance problem, in particular with the nature of the drag onset as the disturbance velocity crosses $c_{\rm min}$ \cite{Raphael1996,Shliomis1997,Sun2001,Browaeys2001,Burghelea2002,Benzaquen2011}. On the other hand, the effect of the finite size of the disturbance on the far-field wake angle at small Bond number (strong capillary effect) has not been described.

The aim of this paper is to characterize the wake behind a moving disturbance
of size of order of the capillary length $\lambda_c$, focusing on the far-field angle of maximum wave amplitude. A series of experiments has been performed using surface-piercing vertical cylinders (periscopes) of various diameters and large immersion depth translated at constant velocity. Using a two-dimensional geometry for the disturbance eliminates the dependence with respect to the immersion depth, which would have varied with velocity for a three-dimensional partially immersed body.  The wake angle here is therefore uniquely determined by the Froude and Bond numbers based on the cylinder diameter.

Our observations suggest that the most remarkable effect of capillarity on the wake geometry is the presence of a capillary-gravity cusp angle (distinct from the usual gravity cusp angle at $\alpha_K \simeq 19.47^\mathrm{o}$), which is related to the minimum of the group velocity at $c_{\rm g,min} \simeq 0.77 c_{\rm min}$ (17~cm~s$^{-1}$ for the air-water interface).  We find that, for Bond number of order unity, the angle of maximum wave amplitude is governed by this capillary-gravity cusp angle, yielding at large velocity
\begin{equation}
\alpha \simeq \frac{c_{\rm g,min}}{U}.
\label{eq:amcgm}
\end{equation}
This law is similar to Eq.~(\ref{eq:am}), although here the effective ``sound velocity'' $c_g$ has a different physical content.   This is because if this disturbance size is not too far from the wavelength of the capillary cusp ($\simeq 2.54 \lambda_c \simeq 4$~cm), then the energy radiated by the disturbance accumulates along this capillary cusp angle. On the other hand, for larger disturbance the wake angle is governed by the pure gravity waves and the law (\ref{eq:am}) is recovered.

The two simple scaling laws (\ref{eq:am}) and (\ref{eq:amcgm}) are derived under the strong assumption that the disturbance is characterized by a single length scale, as is the case for a moving Gaussian pressure disturbance of prescribed size.  In the case of a moving solid body the relation between the disturbance 
size and the resulting pressure distribution depends on the shape of the body, and also on various flow phenomena such as detached boundary layers, wave breaking,  vortex shedding, and so on. In spite of these limitations, the present measurements are well described by Eqs.~(\ref{eq:am}) and (\ref{eq:amcgm}), provided that the effective size of the pressure disturbance is chosen of order of a few cylinder diameters. This is in contrast with streamlined bodies such as ships, for which the effective pressure disturbance at large velocity has essentially the size of the ship.

\section{Experiments}
\label{sec:setup}

\subsection{Experimental setups}

\begin{figure}
\centerline{\includegraphics[width=0.85\linewidth]{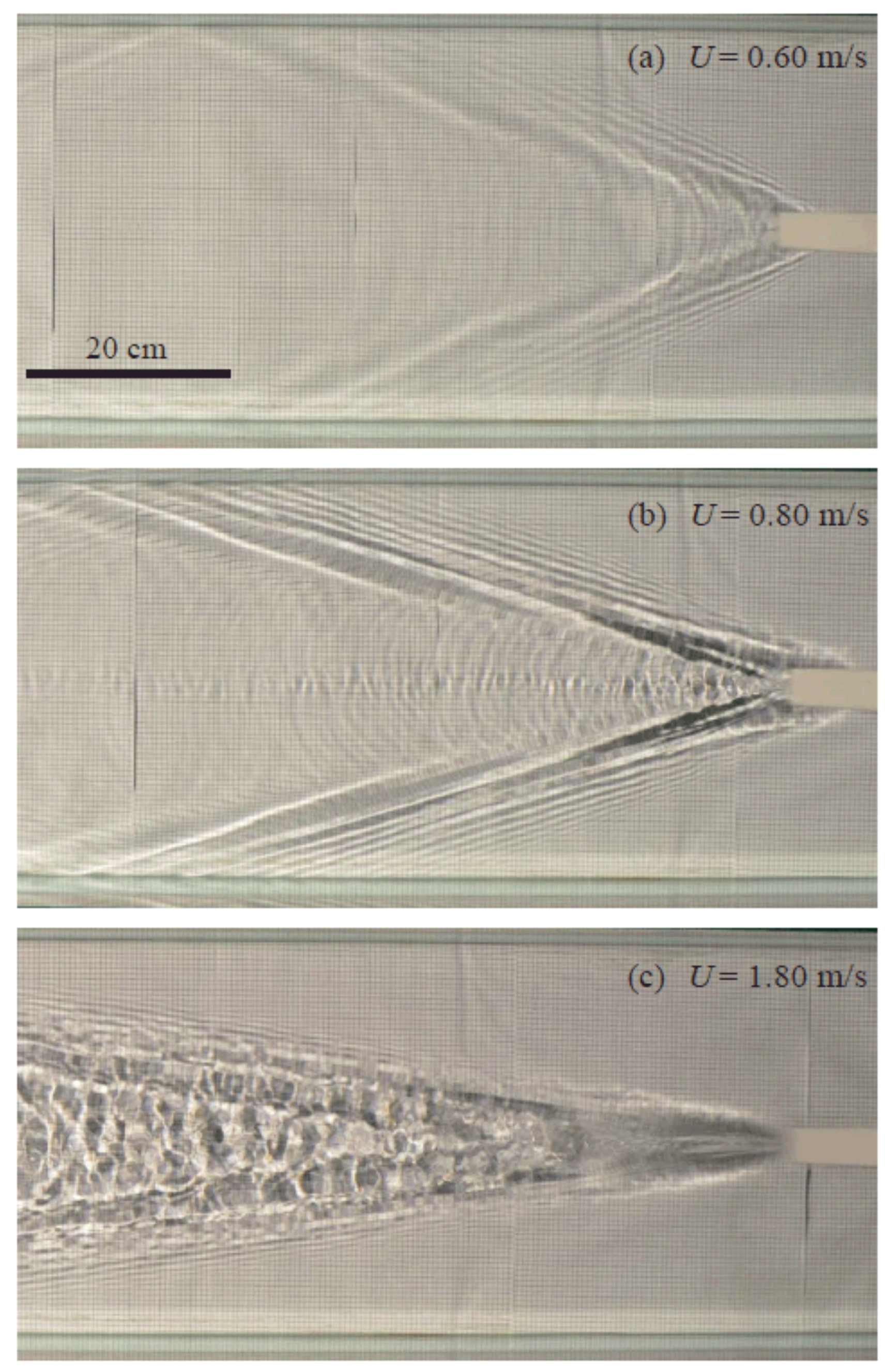}}
\caption{(Color online) Wake patterns in the small-scale towing tank experiments, for a cylinder of diameter $D = 1.5$~mm (Bond number $\mathrm{Bo}_D=0.10$), at velocity $U = 0.60, 0.80$ and $1.80$~m~s$^{-1}$. The tank is 2~m long, and only the last 0.85~m are shown. The waves are visualized by shadowgraphy on the bottom of the water tank.
\label{fig:fast}}
\end{figure}

\begin{figure}
\centerline{\includegraphics[width=0.95\linewidth]{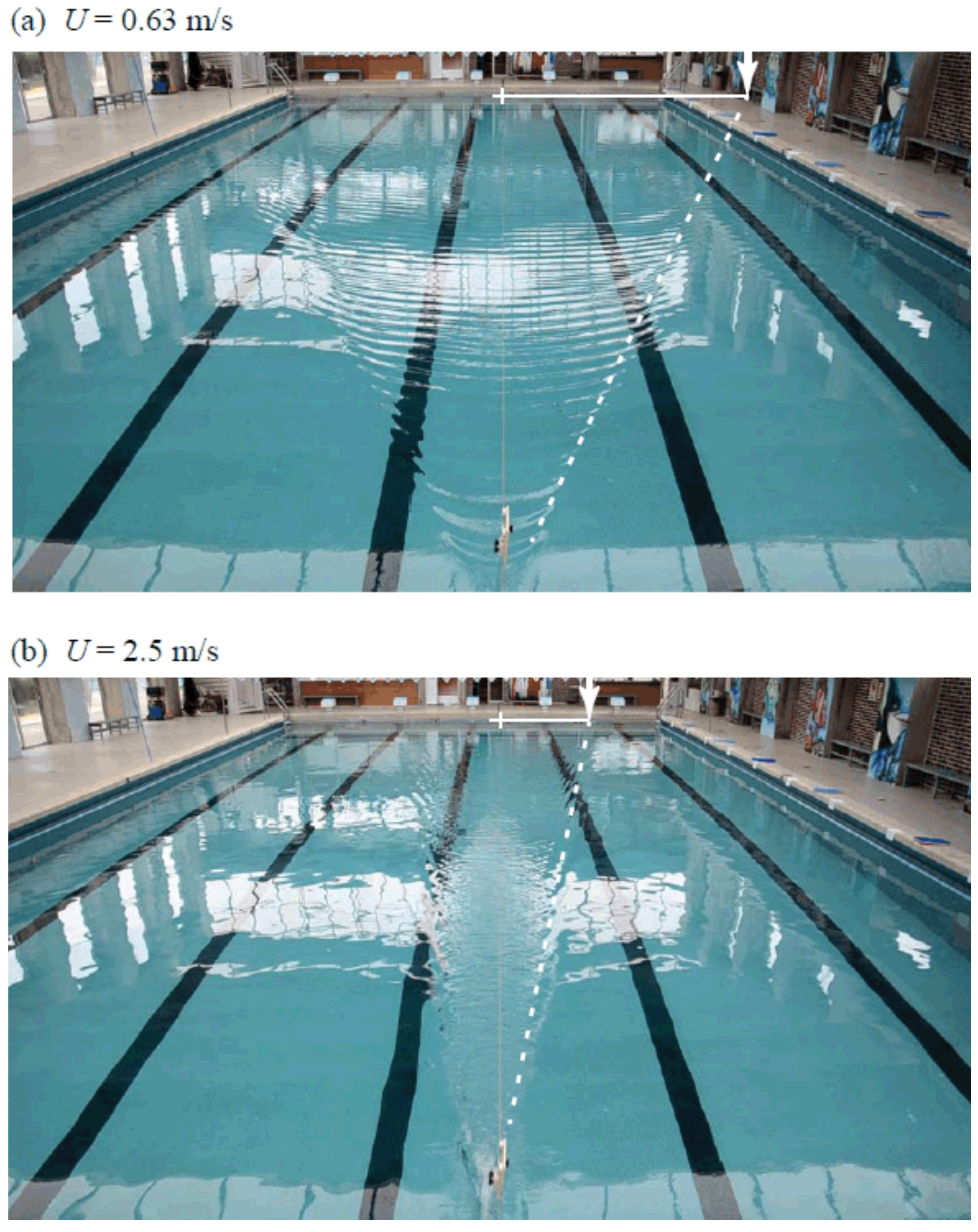}}
\caption{(Color online) Wake patterns in the swimming pool experiments, for a cylinder of diameter $D = 16$~mm (Bond number $\mathrm{Bo}_D=1.0$), at velocity $U = 0.63$ and $2.5$~m~s$^{-1}$. The wake angle is determined from the intersection (shown by the vertical arrow) of the dashed line, going through the waves of maximum amplitude, to the back edge of the pool.
\label{fig:pool}}
\end{figure}

The experiments consists in translating a vertical cylinder, partially immersed in water, at constant velocity, and imaging the resulting wake to measure the angle of maximum wave amplitude. Two series of experiments have been carried out: one in a small-scale towing tank for small cylinder diameters (Fig.~\ref{fig:fast}), and the other in a swimming-pool for larger diameters (Fig.~\ref{fig:pool}). 

The small-scale towing tank is 2~m long and 0.4~m wide and filled to a depth of 0.2~m of tap water. The cylinders are 30-cm-long stainless steel rods, of diameter $D=1.5$ and 5~mm, with at least 10~cm immersed under the water surface. They are hung on a horizontal translation stage driven by a servo-control constant current motor.
For each cylinder, a series of typically 20 runs at different translation velocities have been performed, with $U$ ranging from $c_{\rm min}$ to 3~m~s$^{-1}$ (${\cal U} = U/c_{\rm min}$ between 1 and 13). The acceleration of the translation stage is set between 1 and 10~m~s$^{-2}$ depending on the desired cylinder velocity, so the acceleration length is less than 25\% of the tank length even at the largest velocity.

The swimming pool is 25~m long, 12.5~m wide, and 2~m deep. The cylinders are hollow plastic tubes, 60~cm long, to at least 20~cm of immersion depth, and diameters $D=10,30,$ and 62~mm.  The cylinder is mounted on a carriage with pulley wheels, which is translated along a stainless steel wire rope stretched across the pool. The carriage is pulled by a thread winded on a spool driven by a motor at one end of the pool. While the cylinders remained strictly vertical in the small-scale towing tank, they were slightly sloped backward at high velocities in the swimming-pool experiments because of the strong drag (for the largest diameter the angle relative to the vertical remains less than 5$^{\rm o}$ up to 2~m~s$^{-1}$, but it reaches 20$^{\rm o}$ at 3~m~s$^{-1}$).

In the small-scale experiment, the surface tension of water has been estimated {\it in situ} from the measured wavelengths using the following procedure. The cylinder is towed at a constant velocity $U$ chosen just above the minimum velocity $c_{\rm min}$ of wake onset, for which the gravity wave behind and the capillary wave before the cylinder have nearly the same wavelength. From the measurement of these two wavelengths, the velocity ratio ${\cal U} = U / c_{\rm min}$ can be computed [using Eq.~(\ref{eq:k12}), see next section], from which we deduce the minimum phase velocity $c_{\rm min}$. We found $c_{\rm min} = 21.6\pm 0.2$~cm~s$^{-1}$, which corresponds to a surface tension $\gamma \simeq 55 \pm 2$~mN~m$^{-1}$ and a capillary length $\lambda_c = 2 \pi (\gamma / \rho g)^{1/2} = 14.9 \pm 0.3$~mm.
The surface tension of water in the swimming pool has been measured using a Wilhelmy plate tensiometer, yielding $\gamma \simeq 66 \pm 3$~mN~m$^{-1}$ and hence
$c_{\rm min} = 22.6\pm 0.3$~cm~s$^{-1}$ and $\lambda_c = 16.3 \pm ~ 0.4$~mm.
For the range of cylinder diameters used here, the Bond number $\mathrm{Bo}_D = D /\lambda_c$ ranges from 0.1 to 4.2, with a precision of $\pm 5\%$. The maximum wavelength excited by the disturbance being of order of the cylinder diameter, which is comfortably smaller than the water depth in both setups, the wakes can be considered in the deep water regime. The Reynolds number $Re = UD/\nu$ covers a wide range, from 350 to 180~000, for which the hydrodynamic wake is unstationary to fully turbulent. 

During each run movies were taken using a digital camera. The camera was located above the tank for the small-scale experiments, and at one end of the wire rope at a height of 3~m above the water surface for the swimming-pool experiments. The images are analyzed in the second half of the cylinder course, well after the acceleration phase, so the wake is well developed. For each image, the wake arms are defined from the most visible waves, i.e. from the waves showing the most contrasted light pattern. For the small-scale experiments, the waves appear in the form of dark and bright stripes on the bottom of the water tank (shadowgraphy), while for the swimming-pool experiments they are visualized by reflection of natural light. The biases introduced by the different visualization methods are discussed in the appendix. For the swimming-pool experiments, the wake angle is determined by drawing lines going through the waves of larger amplitude and extended to the back edge of the pool (see dashed lines in Fig.~\ref{fig:pool}).  The uncertainty is $\pm 1^{\rm o}$ for the small-scale experiments and $\pm 2^{\rm o}$ for the swimming-pool experiments.

\begin{figure}
\centerline{\includegraphics[width=0.90\linewidth]{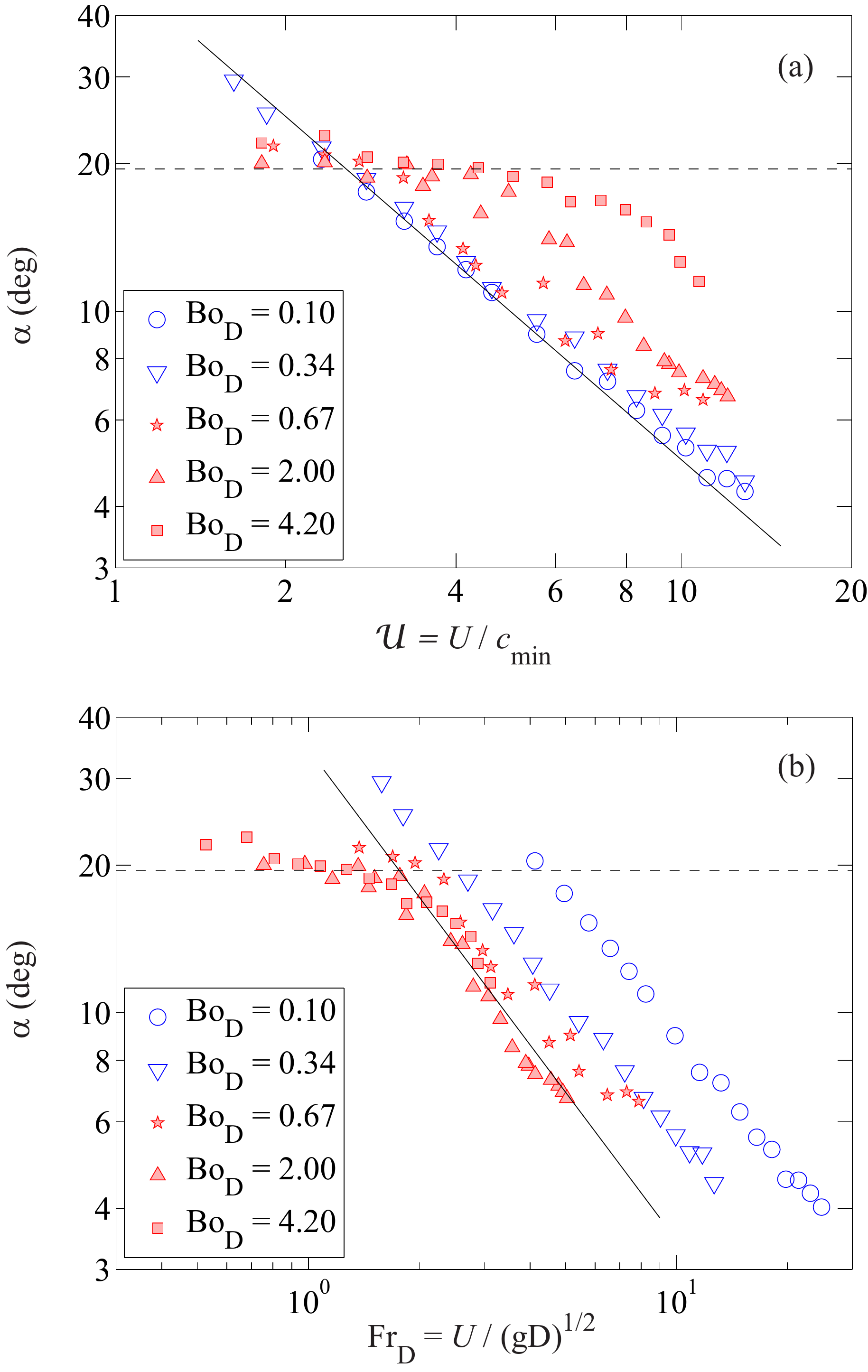}}
\caption{(Color online) Wake angle as a function of the cylinder velocity $U$, normalized by $c_{\rm min}$ in (a), and by $\sqrt{gD}$ in (b). Open symbols (blue): Small-scale experiments; filled symbols (red): swimming-pool experiments. The solid lines show best fits of the data at large velocity: (a) $\alpha = 0.85 \, c_{\rm min}/U$ for $\mathrm{Bo}_D = 0.10$ and 0.34 and (b) $\alpha = 0.5 / \mathrm{Fr}_D$ for $\mathrm{Bo}_D = 2.0$ and 4.2 (numerical values are given for $\alpha$ in radians).
\label{fig:aexpFr}}
\end{figure}

\subsection{Experimental wake angles}
\label{sec:ewa}

The wake angles measured in the two setups for the various cylinder diameters, plotted in Fig.~\ref{fig:aexpFr}, display a systematic decrease as $U^{-1}$ at large velocity. In order to discreminate the wakes dominated by capillary or gravity effects,  two normalizations are used: ${\cal U} = U / c_{\rm min}$ in Fig.~\ref{fig:aexpFr}(a) and $\mathrm{Fr}_D = U / \sqrt{gD}$ in Fig.~\ref{fig:aexpFr}(b).

The normalization ${\cal U} = U / c_{\rm min}$ in Fig.~\ref{fig:aexpFr}(a) provides a good collapse of the wake angles at small Bond numbers, essentially for the two data sets from the small-scale experiments ($\mathrm{Bo}_D = 0.10$ and 0.34) but also to some extent for the smaller cylinder diameter in the swimming-pool experiments ($\mathrm{Bo}_D = 0.67$). No evidence of Kelvin wake angle $\alpha_K = 19.47^\mathrm{o}$ is found for the smaller Bond numbers, but rather a continuous decrease from $\alpha \simeq 30^\mathrm{o}$ to $\sim 4^\mathrm{o}$. A best fit of the first two data sets gives $\alpha \simeq 0.85 \, c_{\rm min}/U$ (the prefactor is given for $\alpha$ in radians).

The normalization $\mathrm{Fr}_D = U / \sqrt{gD}$ in Fig.~\ref{fig:aexpFr}(b) provides a better collapse at larger Bond numbers, essentially for the two largest cylinder diameters in the swimming-pool experiments ($\mathrm{Bo}_D = 2.0$ and 4.2). Here, a clear transition is found between the Kelvin angle, for $\mathrm{Fr}_D < 1.5$, and a Mach-like regime consistent with $\alpha \simeq a / \mathrm{Fr}_D$ at larger Froude number. A best fit of the last two data sets for $\mathrm{Fr}_D > 2$ gives $a \simeq 0.5 \pm 0.1$.

Interestingly, the value $a \simeq 0.5$ is significantly larger than the one found for  ship wakes, $a \simeq 0.2$. According to the analysis of Ref.~\cite{Rabaud2013}, this prefactor is expected to scale as $(\lambda_f/D)^{1/2}$, where $\lambda_f$ is the dominant wave length excited by the disturbance. Its value depends on the shape of the disturbance, and on the detail of the flow around it. The detached flow around bluff bodies such as cylinders implies a disturbed region significantly larger than the body, and hence a value of $a$ larger than for streamlined bodies such as ships. Comparing the values of $a$ for ships and cylinders suggests that a ship of length $L$ primarily excites  waves of wavelength $\lambda_f \simeq L$, whereas we have $\lambda_f \simeq (6 \pm 2) D$ here, yielding $a \simeq 0.2 (\lambda_f/D)^{1/2} \simeq 0.5$. Consistently, the transition Froude number between the Kelvin and the Mach regimes is also shifted: one has $Fr_L \simeq 0.6$ for ships, and $\mathrm{Fr}_D \simeq 0.6 (\lambda_f/D)^{1/2} \simeq 1.5$ here.

\section{Capillary-gravity crest lines}
\label{sec:crests}

In order to model the far-field wake angle of a disturbance of finite size, it is necessary to describe first the geometry of the crest lines, which provides the skeleton of the capillary-gravity wake pattern, without specifying at this point where energy radiated by the disturbance is actually located on this skeleton. The reader is referred to Refs.~\cite{Lamb,Crapper1964,Binnie1965,Yih2,Doyle2013} for a complete description of the capillary-gravity crest line pattern, which we briefly summarize here for convenience.  We give special emphasis on the gravity and capillary cusp angles, which are of first importance when examining the effects of the finite size of the disturbance (Sec.~\ref{sec:fse}).

\begin{figure}
\centerline{\includegraphics[width=0.90\linewidth]{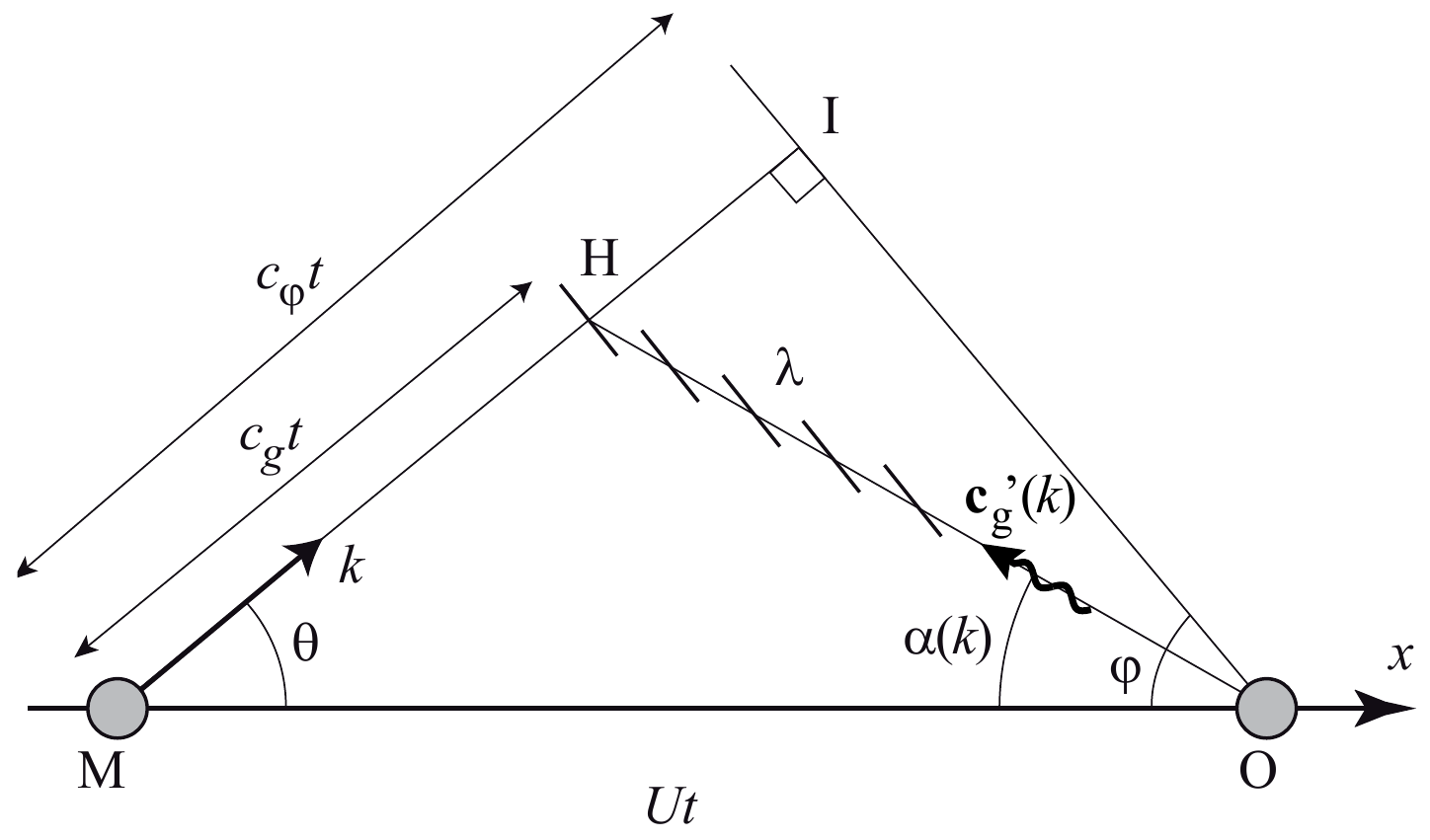}}
\caption{Construction of the stationary wake pattern. The disturbance is at point O at time $0$, and we consider a wave vector ${\bf k}$ emitted at point M at time $-t$. In the frame of the disturbance, its energy propagates along the radiation angle $\alpha(k)$. This is the direction of the relative group velocity ${\bf c}'_g(k) = {\bf c}_g(k)  - {\bf U}$. 
\label{fig:sketch}}
\end{figure}

\begin{figure}
\centerline{\includegraphics[width=0.9\linewidth]{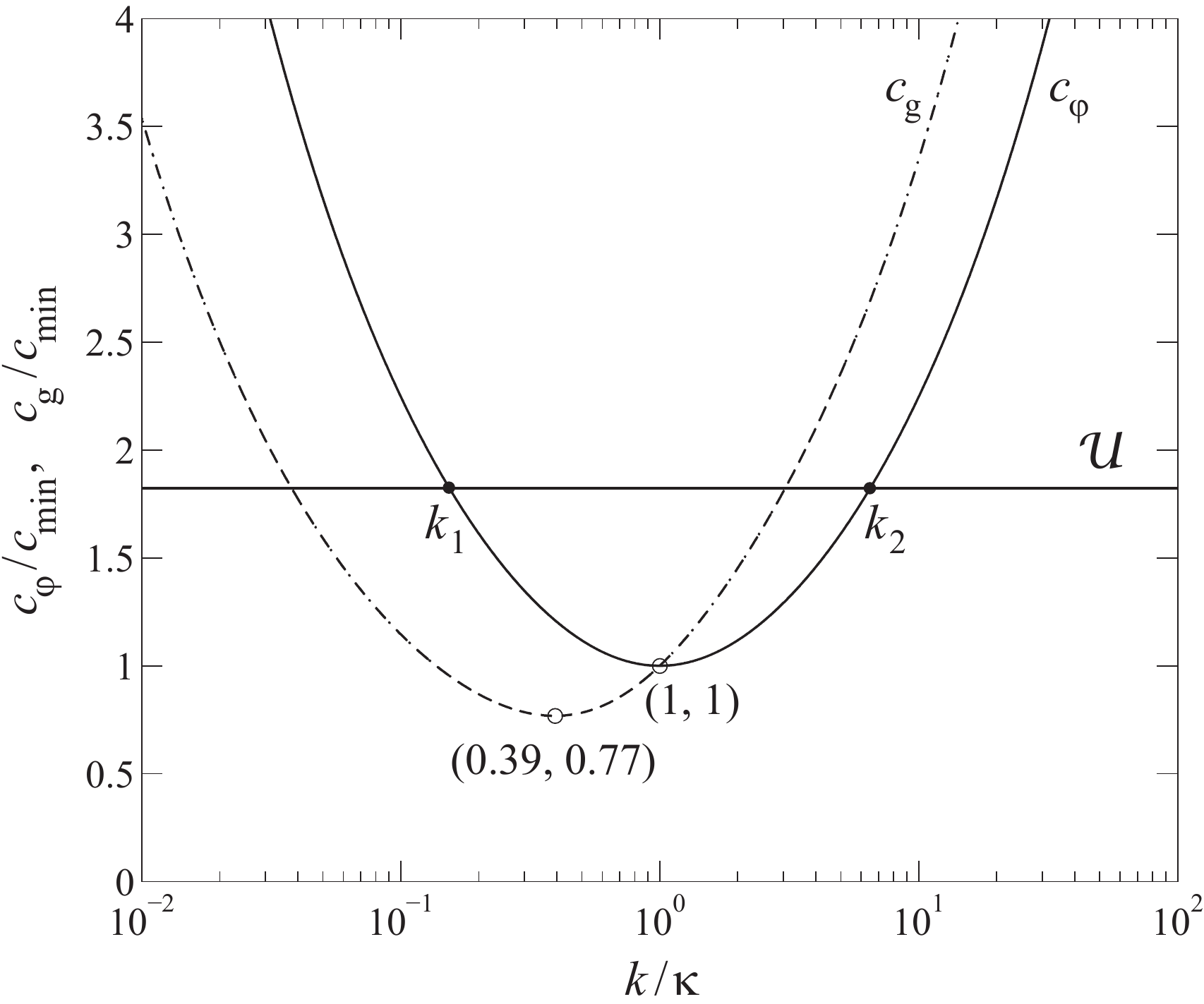}}
\caption{Phase velocity $c_\varphi$ and group velocity $c_g$, normalized by the minimum phase velocity $c_{\rm min}$, as a function of the normalized wavenumber $k/\kappa$, with $\kappa = (\rho g / \gamma)^{1/2}$. The minimum group velocity is $c_{\rm g,min} / c_{\rm min} \simeq 0.77$, at $k/\kappa \simeq 0.39$. A disturbance velocity ${\cal U} = U / c_{\rm min} > 1$ selects a range $[k_1, k_2]$ such that the stationary condition (\ref{eq:sc}) is satisfied. 
\label{fig:cphi}}
\end{figure}

We consider the stationary phase wake pattern generated by a disturbance moving
at constant velocity $U$ in the $x$ direction (Fig.~\ref{fig:sketch}).
For any wave vector ${\bf k}$ emitted from point M at time $-t$,
the condition of stationarity with respect to the disturbance in O implies 
\begin{equation}
U \cos \theta = c_\varphi(k),
\label{eq:sc}
\end{equation}
where $\theta$ is the angle between ${\bf k}$ and ${\bf U}$ and $c_\varphi(k) = \omega(k)/k$ is
the phase velocity. Equation~(\ref{eq:sc}) is the statement that the relative frequency
in the moving frame, $\Omega(\bf k) = \omega({\bf k}) - {\bf U} \cdot {\bf k}$,
is zero. The frequency is given by the dispersion
relation for capillary-gravity waves in deep water, $\omega (k) = ( gk + \gamma k^3/\rho )^{1/2}$,
with $\rho$ the density, $\gamma$ the surface tension and $g$ the gravity.
The phase velocity (plotted in Fig.~\ref{fig:cphi}) has a minimum, equal to $c_{\rm min} = (4 g \gamma / \rho)^{1/4}$, at the capillary-gravity wavenumber $\kappa = (\rho g / \gamma)^{1/2}$.
The stationary condition (\ref{eq:sc}) therefore can be satisfied only for 
$U \geq c_{\rm min}$. For a given velocity ratio ${\cal U} = U /  c_{\rm min} > 1$, there is a range of wave numbers $k \in [k_1, k_2]$ satisfying Eq.~(\ref{eq:sc}), such that
\begin{equation}
\frac{k_{1,2}}{\kappa} = {\cal U}^2 \mp ( {\cal U}^4 - 1 )^{1/2}.
\label{eq:k12}
\end{equation}
For ${\cal U} \simeq 1$, one has $k_1 \simeq k_2 \simeq \kappa$. On the other hand, for ${\cal U} \gg 1$, $k_1$ tends to the pure gravity wavenumber $k_g = g/U^2$, and $k_2$ tends to the pure capillary wavenumber $k_c = \rho U^2/\gamma$, so the range of wavenumbers satisfying (\ref{eq:sc}) rapidly grows as $k_2 / k_1 \simeq 4 {\cal U}^4$.

Each wavenumber $k \in [k_1, k_2]$ contributing to the stationary pattern is associated to
a {\it radiation angle} $\alpha(k)$, i.e. an angle at which the energy radiated from the disturbance propagates. This is the angle between $-{\bf U}$ and the group velocity in the moving frame, given by ${\bf c}'_g = \nabla_{\bf k} \Omega = {\bf c}_g - {\bf U}$, where ${\bf c}_g = \nabla_{\bf k} \omega$ is the group velocity in the frame of the liquid at rest, and $\nabla_{\bf k}$ denotes the gradient in the Fourier space. The general derivation of $\alpha(k)$ for arbitrary dispersion relation can be found in Refs.~\cite{Ursell1960,Keller}. We briefly recall here this derivation, following the geometrical approach of Crawford \cite{Crawford1984} extended to the case of capillary-gravity waves (see Carusotto and Rousseaux \cite{Rousseaux2013} and Doyle and McKenzie \cite{Doyle2013} for a similar derivation in the Fourier space).

We consider in Fig.~\ref{fig:sketch} a wave of wavenumber ${\bf k}$ emitted from a point M at time $-t$ satisfying the stationary condition (\ref{eq:sc}). At time $0$, the phase of the wave reaches the point I, with MI $= c_\varphi t$, such that MI $\perp$ OI. Denoting $\varphi$ the angle between OM and OI, one has $\tan \varphi =$ MI/OI$ = c_\varphi / \sqrt{U^2 - c_\varphi^2}$.  Since the energy emitted from M at time $-t$ travels at the group velocity $c_g = \partial \omega / \partial k$, it reaches the point H, with MH $= c_g t$. For $k < \kappa$ (gravity waves), one has $c_g < c_\varphi$, so the wave packet in H does not reach the point I, whereas for $k > \kappa$ (capillary waves) the wave packet travels beyond I. In the limit case of pure gravity waves, one has $c_g = c_\varphi / 2$, so H is the middle of MI.

\begin{figure}
\centerline{\includegraphics[width=0.9\linewidth]{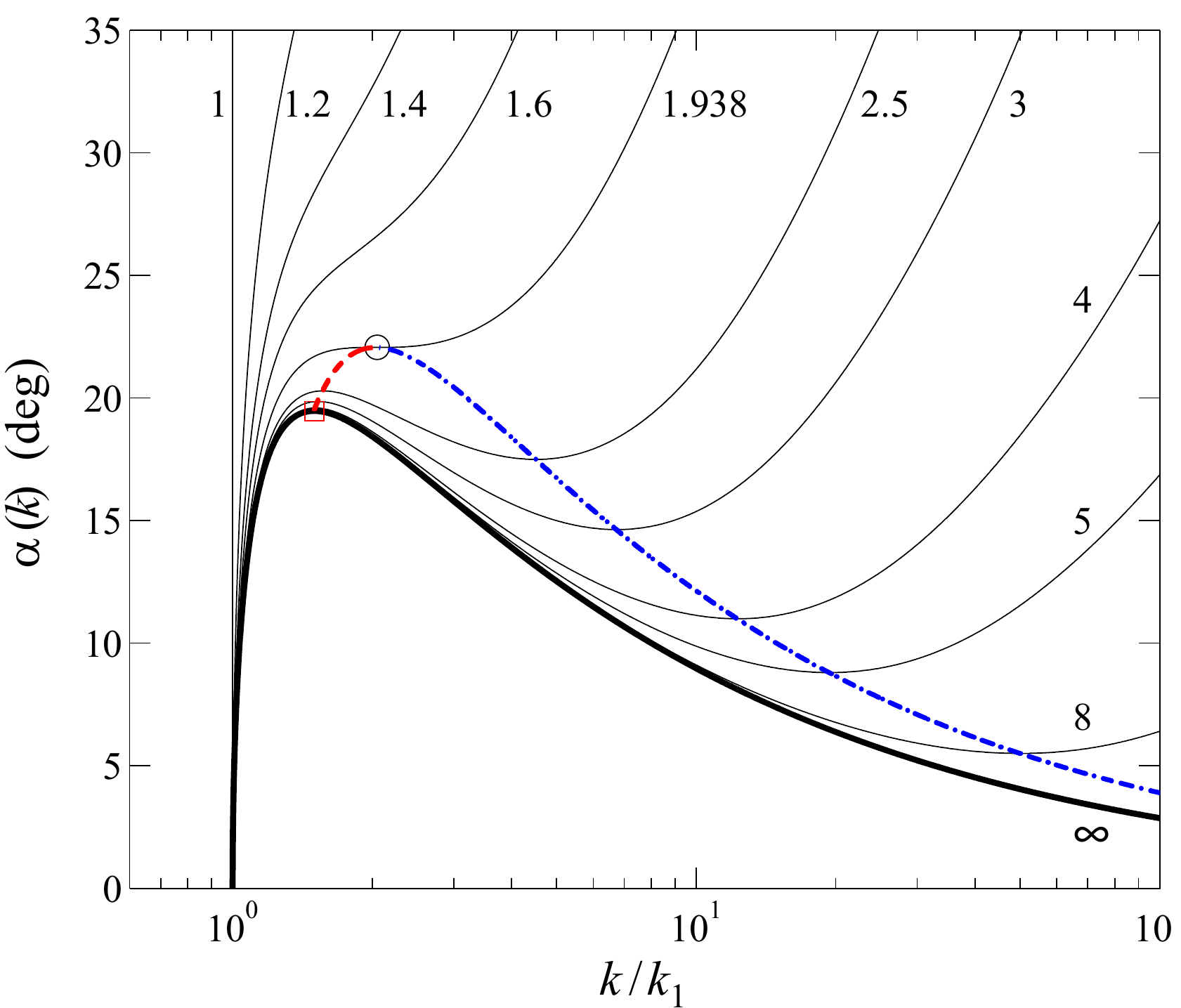}}
\caption{(Color online) Radiation angle $\alpha(k)$ as a function of the wavenumber $k$ normalized by
the minimum wavenumber $k_1$ given by Eq.~(\ref{eq:k12}). The bold curve shows the pure gravity case (${\cal U} \rightarrow \infty$), and the thin curves show capillary-gravity cases for ${\cal U}$ between 1 and 8. The dashed lines show the locus of the upper and lower extrema of $\alpha(k)$, corresponding to the gravity and capillary cusp angles, $\alpha_g^{\rm cusp}$ ($--$, red) and $\alpha_c^{\rm cusp}$ ($-\cdot-$, blue). $\circ$~: Onset of existence of the two cusps for ${\cal U}^* = 1.938$, at $\alpha^* \simeq 22.06^\mathrm{o}$. $\square$~: Asymptotic gravity cusp angle for ${\cal U} \gg 1$, which corresponds to the Kelvin angle $\sin^{-1} (1/3) \simeq 19.47^\mathrm{o}$.
\label{fig:akc}}
\end{figure}

The direction OH defines the radiation angle $\alpha(k)$ at which the energy of a given wavenumber $k$ emitted from all points between M and O and satisfying the stationary condition is located. Using the relation $\tan (\varphi - \alpha) = $ HI/OI $ = (c_\varphi - c_g) / \sqrt{U^2 - c_\varphi^2}$ finally yields the following~\cite{Ursell1960,Keller}:
\begin{equation}
\tan \alpha(k) = \frac{c_g(k)\sqrt{U^2 - c_\varphi^2(k)}}{U^2 - c_g(k) c_\varphi(k)}.
\label{eq:tanak}
\end{equation}
This angle is plotted in Fig.~\ref{fig:akc} for different ratios ${\cal U}= U/c_{\rm min}$. It is defined for $k$ in the interval $[k_1,k_2]$ allowed by the disturbance velocity ${\cal U}$. It satisfies $\alpha(k_1) = 0$, corresponding to the transverse gravity waves radiated behind the disturbance, and $\alpha(k_2) = 180^\mathrm{o}$, corresponding to the transverse capillary waves radiated in the front of the disturbance. 

\begin{figure}
\centerline{\includegraphics[width=0.95\linewidth]{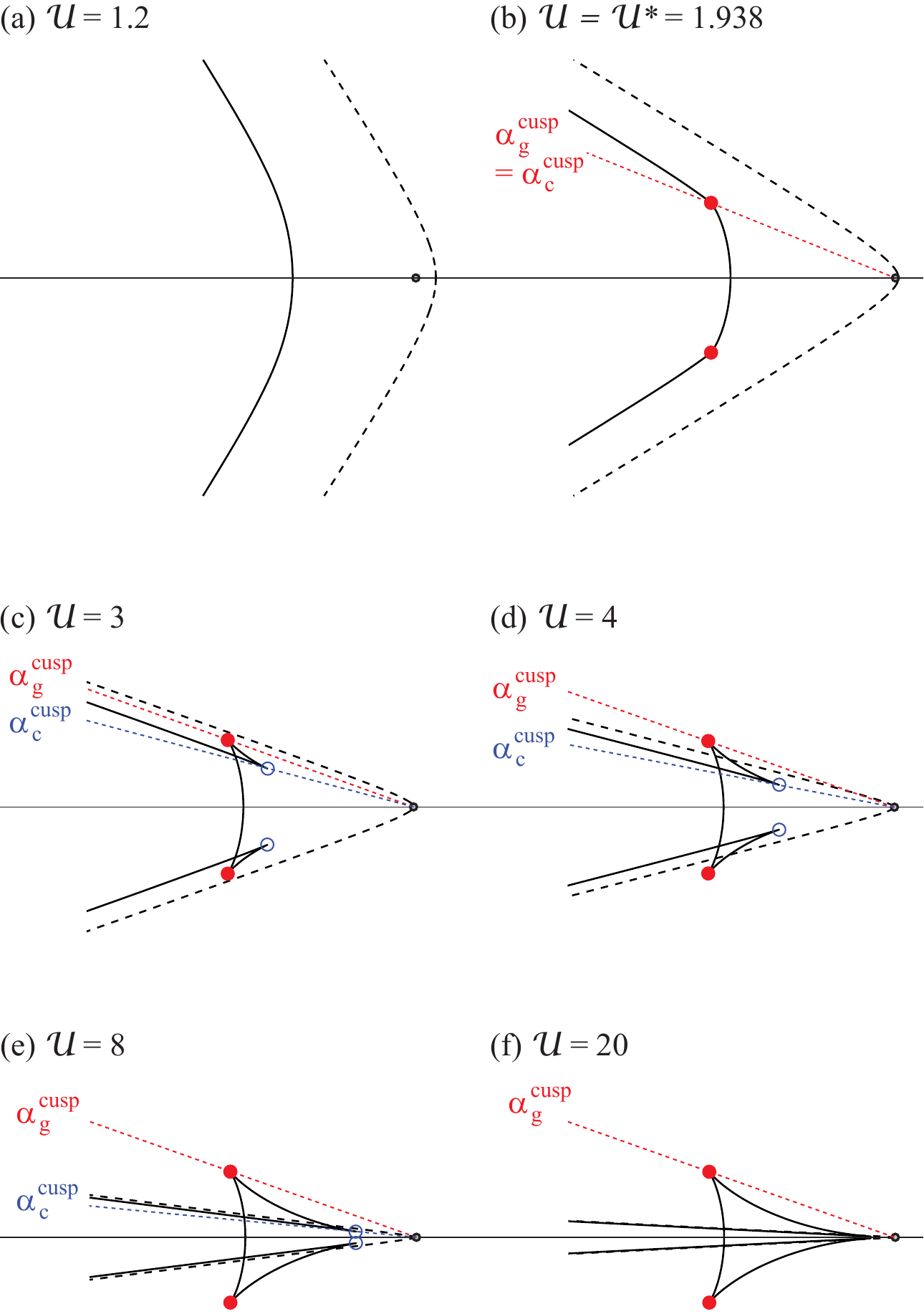}}
\caption{(Color online) Patterns of iso-phase lines (e.g., crest lines) for six velocity ratios ${\cal U}$. For each velocity, only two crest lines are shown, corresponding to the capillary branch (long dashed line) and the gravity branch (continuous line). Note that a phase shift of $\pi/2$ appears at each cusp point. The filled circles (red) show the gravity cusp and the empty circles (blue) show the capillary cusp,
present for ${\cal U} > {\cal U}^* = 1.938$ [the two cusps merge at ${\cal U}^*$ (b)]. The short dashed lines show the cusp angles $\alpha_g^{\rm cusp}$ (red) and $\alpha_c^{\rm cusp}$ (blue), along which the energy radiated from the source accumulates.
\label{fig:cp}}
\end{figure}

For ${\cal U} \rightarrow \infty$, the pure gravity radiation angle 
is recovered~\cite{Rabaud2013}: One has $c_g = c_\varphi/2$, and Eq.~(\ref{eq:tanak}) reduces to
\begin{equation}
\tan \alpha(k) = \frac{\sqrt{k/k_g - 1}}{2 k/k_g - 1},
\label{eq:tanakg}
\end{equation}
with $k_g = k_1 = g/U^2$. This law, plotted as bold line in Fig.~\ref{fig:akc}, shows a single extremum at $\alpha_K = \sin^{-1} (1/3) = 19.47^\mathrm{o}$: this is the classical Kelvin angle, at which the pattern of crest lines show a cusp. For finite ${\cal U} \geq 1$, the radiation angle curve is more complicated, and it is instructive to examine its behavior in relation to the shape of the crest lines, which we plot in Fig.~\ref{fig:cp}~\cite{Lamb,Crapper1964,Binnie1965}. At small disturbance velocity, $\alpha(k)$ is a monotonous function of $k$, increasing from 0 to $180^\mathrm{o}$ in the interval $[k_1, k_2]$, indicating that energy is smoothly radiated all around the disturbance (this radiation tends to be isotropic in the limit ${\cal U} \rightarrow 1$). The resulting smooth and slightly curved capillary ripples in front of the disturbance are known as Poncelet ripples~\cite{Darrigol,Bouasse} [Fig.~\ref{fig:cp}(a)]. As the velocity is increased, $\alpha(k)$ is no longer monotonous, and shows two local extrema. In the vicinity of these extrema, there exists a small range of wavenumbers for which $\alpha(k)$ is locally constant, indicating the formation of two cusps in the crest lines, both located behind the disturbance (see the dashed lines in Fig.~\ref{fig:cp}).  We call them gravity ($\alpha_g^{\rm cusp}$) and capillary ($\alpha_c^{\rm cusp}$) cusp angles --- although the second one actually results from mixed gravity and capillary effects. The locus $(k_{g}^{\rm cusp}, \alpha_{g}^{\rm cusp})$ and $(k_{c}^{\rm cusp}, \alpha_{c}^{\rm cusp})$ are shown as dashed curves in Fig.~\ref{fig:akc}.  These two cusp angles play a major role in the shape of the far-field wake when the finite size of the disturbance is considered (Sec.~\ref{sec:fse}): In the presence of such cusps, the energy of the disturbance is no longer radiated smoothly around the disturbance but rather concentrates along the cusps.

\begin{figure}
\centerline{\includegraphics[width=0.80\linewidth]{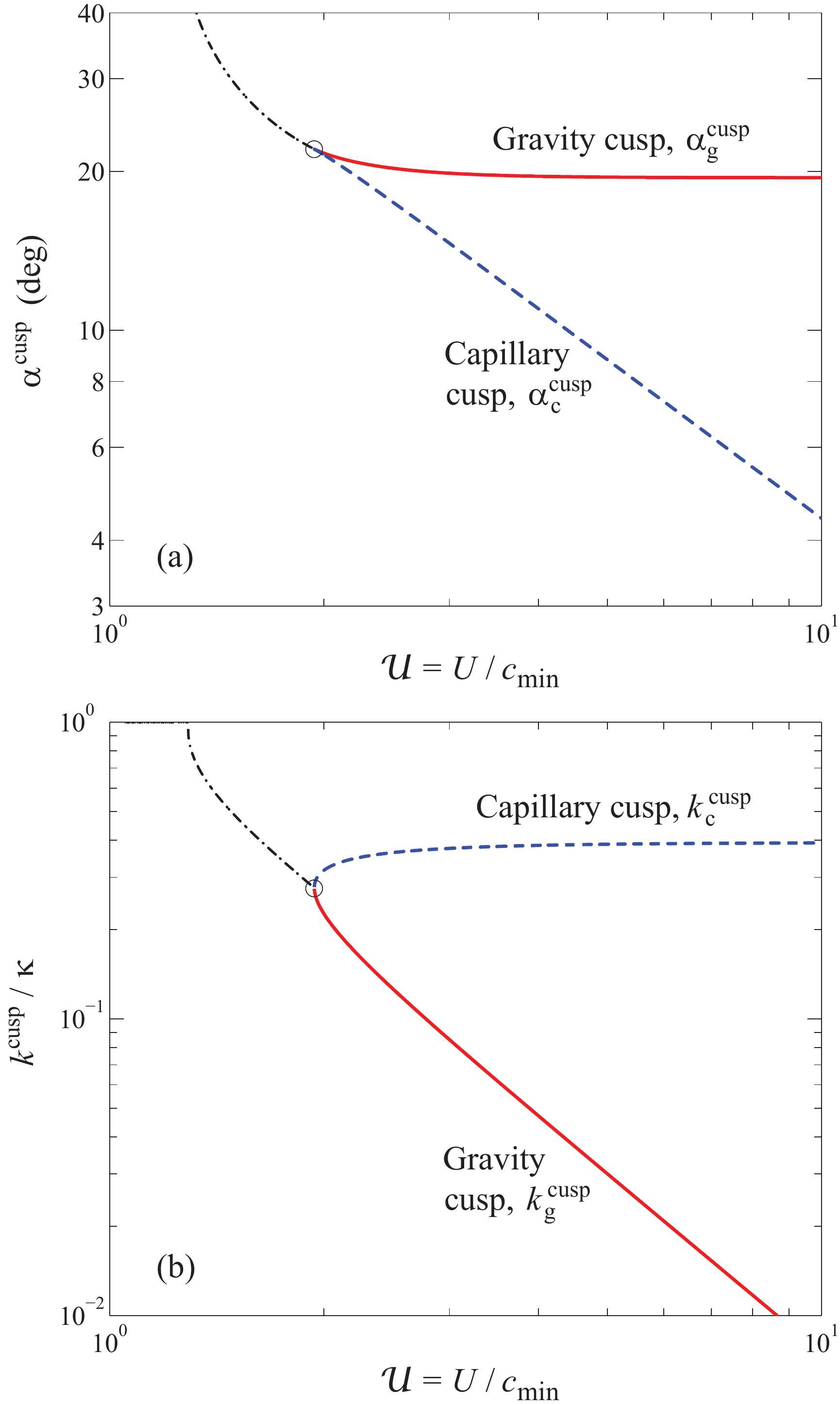}}
\caption{(Color online) (a) Gravity and capillary cusp angles $\alpha_g^{\rm cusp}$ and $\alpha_c^{\rm cusp}$, and (b) normalized cusp wavenumbers $k_g^{\rm cusp}/\kappa$ and $k_c^{\rm cusp}/\kappa$, as a function of the normalized velocity ${\cal U}$. The circle indicates the onset of the cusp at ${\cal U}^* \simeq 1.938$ ($\alpha^* \simeq 22.06^\mathrm{o}$ and $k^*/\kappa \simeq 0.275$).
The line $- \cdot -$ shows the cusp precursor (angle such that $\partial \alpha / \partial k$ is minimum), where a preferential accumulation of energy may take place even before the apparition of the cusps.
\label{fig:cusp}}
\end{figure}

The two cusp angles and the corresponding wavenumbers are plotted in Fig.~\ref{fig:cusp} as a function of the velocity ratio ${\cal U}$. They appear through a saddle-node bifurcation at the velocity ${\cal U}^* \simeq 1.938$, as first noticed by Binnie~\cite{Binnie1965} (see also Refs.~\cite{Yih2,Doyle2013}). Note that the angle corresponding to the minimum of $\partial \alpha / \partial k$ is also of interest: The energy of the disturbance may already accumulate near this cusp precursor even for ${\cal U} < {\cal U}^*$.
At the onset, $\alpha_g^{\rm cusp}$ and $\alpha_c^{\rm cusp}$ are both equal to $\alpha^* \simeq 22.06^\mathrm{o}$ (marked by a symbol $\circ$ in Figs.~\ref{fig:akc} and \ref{fig:cusp}), which is slightly larger than the Kelvin angle $19.47^\mathrm{o}$ for pure gravity waves. At this point the crest lines [Fig.~\ref{fig:cp}(b)] show only a weak change of curvature. As the disturbance velocity is increased, $\alpha_g^{\rm cusp}$ tends rapidly towards the classical Kelvin angle, shown by a symbol $\square$ in Fig.~\ref{fig:akc}, with a departure from the Kelvin angle decreasing as ${\cal U}^{-4}$. On the other hand, the capillary cusp angle $\alpha_c^{\rm cusp}$ is a decreasing function of ${\cal U}$. In the limit ${\cal U} \gg 1$, since the wavenumbers $k$ in the vicinity of the minimum of $\alpha(k)$ satisfy $c_g(k) \ll U$ and $c_\varphi(k) \ll U$, Eq.~(\ref{eq:tanak}) can be approximated by
$$
\alpha(k) \simeq \frac{c_g(k)}{U}.
$$
The capillary cusp angle $\alpha_c^{\rm cusp}$ is therefore found at the wavenumber $k_{c}^{\rm cusp}$ satisfying $\partial \alpha / \partial k = U^{-1} \partial c_g / \partial k = 0$, i.e. at the minimum group velocity $c_{\rm g,min}$ (see Fig.~\ref{fig:cphi}). The capillary cusp angle is therefore given by
\begin{equation}
\alpha_c^{\rm cusp} \simeq \frac{c_{\rm g,min}}{U} = \frac{A}{{\cal U}},
\label{eq:acusp2}
\end{equation}
where $A = c_{\rm g,min} / c_{\rm min} = \frac{1}{2} 3^{3/8} (\sqrt{3}-1) (2-\sqrt{3})^{-1/4} \simeq 0.768$. Figure~\ref{fig:cusp}(a) shows that the law (\ref{eq:acusp2}) turns out to hold even very close to the onset of the cusp ${\cal U}^*$.  The wavenumber of this minimum group velocity, $k_{c}^{\rm cusp} / \kappa = \sqrt{2/\sqrt{3}-1} \simeq 0.393$, is in the gravity branch of the dispersion relation, indicating that capillary effects may be important even for disturbance of size significantly larger than the capillary length (in practice for size $2.54 \lambda_c \simeq 40$~mm for the air-water interface).

\section{Angle of maximum wave amplitude}
\label{sec:fse}

\subsection{Modeling of the finite size effects of the disturbance}

We now consider the influence of the disturbance size on the angular distribution of energy in the far-field wake, with the assumption that the waves remain linear in this problem. Although the pattern of crest lines itself is not affected by the disturbance size, different regions of the pattern receive different amounts of energy depending on the spectrum of the disturbance, which has a strong impact on the overall shape of the surface elevation pattern.

We model in the following the disturbance as a pressure distribution and make use of the key result of the Cauchy-Poisson initial value problem~\cite{Havelock1908,Lamb,Wehausen,Whitham,Lighthill}: The waves of larger amplitude generated by an initial pressure disturbance of characteristic size $L$ are contained in a wavepacket traveling at the group velocity selected by the size $L$. For instance, in the case of pure gravity waves [of group velocity $c_g(k) = \frac{1}{2} \sqrt{g/k}$], although wavelengths much larger than $L$ may be excited by the disturbance, they travel much faster than the wavelengths of order of $L$, so their energy is stretched over large distances, and their amplitude decreases accordingly. Let us consider the axisymmetric wave dispersion originating from an initial surface elevation $\zeta_0(r)$ at $t=0$. Using the method of stationary phase, the envelope of the wave train for arbitrary dispersion relation can be written as~\cite{Lamb}
\begin{equation}
\zeta(r,t) \sim \hat \zeta(k=k_0(r,t)) \left(r^2 t \Big| \frac{\partial c_g}{\partial k} (k=k_0(r,t)) \Big| \right)^{-1/2}
\label{eq:etaenv}
\end{equation}
[provided that $\partial c_g /\partial k (k_0) \neq 0$], where $\hat \zeta(k)$ is the Fourier transform of $\zeta_0(r)$, and $k_0(r,t)$ is the local wave number satisfying $c_g (k_0) = r/t$. We consider here for simplicity pure gravity waves excited by an initial Gaussian surface elevation~\cite{prefactor} given by $\zeta_0(r) = h \exp[-\pi^2(r/L)^2]$, of Fourier transform $\hat \zeta(k) \propto h \exp[-(kL)^2/4\pi^2]$. Solving for $c_g (k_0) = r/t$ yields $k_0 (r,t) = g t^2 / 4 r^2$, from which the maximum of the wave envelope (\ref{eq:etaenv}) at given time $t$ is found at $r_{max} (t) = C t$, with $C = a\sqrt{gL}$ and $a=1 / (40^{1/4} \pi^{1/2})\simeq 0.22$. If we consider now the wake problem as a succession of such wave trains emitted by a moving surface elevation at velocity $U$, the resulting stationary pattern has maximum energy approximately at $\sin \alpha \simeq C/U = a/\mathrm{Fr}$ with $\mathrm{Fr} = U/\sqrt{gL}$, in agreement with Ref.~\cite{Darmon2014}. The local wave number $k_0(r,t)$ in the center of the wave packet [at $r \simeq r_{max}(t)$] is given in this case by $\lambda_f = 8 \pi a^2 L = \sqrt{2} L$, confirming that the wavelength of maximum amplitude in the wave packet is of order of the disturbance size \cite{prefactor2}.

The generalization of this simplified approach to the capillary-gravity case is complicated by the fact that $c_g (k_0) = r/t$ has now two solutions, one on the capillary branch and one on the gravity branch, resulting in two superimposed wave packets \cite{Whitham}. One can, however, proceed qualitatively as follows. Since the wave packet radiated by the disturbance of size $L$ is composed near its maximum of wave numbers of order of $k_f \simeq L^{-1}$, it may be characterized by an effective spectrum centered around $k \simeq k_f$. The wave packet being localized in space, the typical width $\Delta k$ of the energy-containing wave number range is also of order of $k_f$. In the frame of the disturbance, the energy of each wave number $k$ is radiated along the direction given by the radiation angle $\alpha(k)$ shown in Fig.~\ref{fig:akc}. As a consequence, the dominant wave number $k_f$ is radiated along the angle $\alpha(k_f)$ and, except in the vicinity of one of the two cusp angles, this radiation takes place within an angular aperture $\Delta \alpha \simeq |\partial \alpha / \partial k| \Delta k$. On the other hand, if a significant amount of energy is radiated in the vicinity of a cusp wave number, i.e., if either $k_g^{\rm cusp}$ or  $k_c^{\rm cusp}$ fall in the energy-containing range $\Delta k$, the wake angle is given by the corresponding cusp angle, which concentrates most of the energy radiated by the disturbance.

\begin{figure}
\centerline{\includegraphics[width=0.8\linewidth]{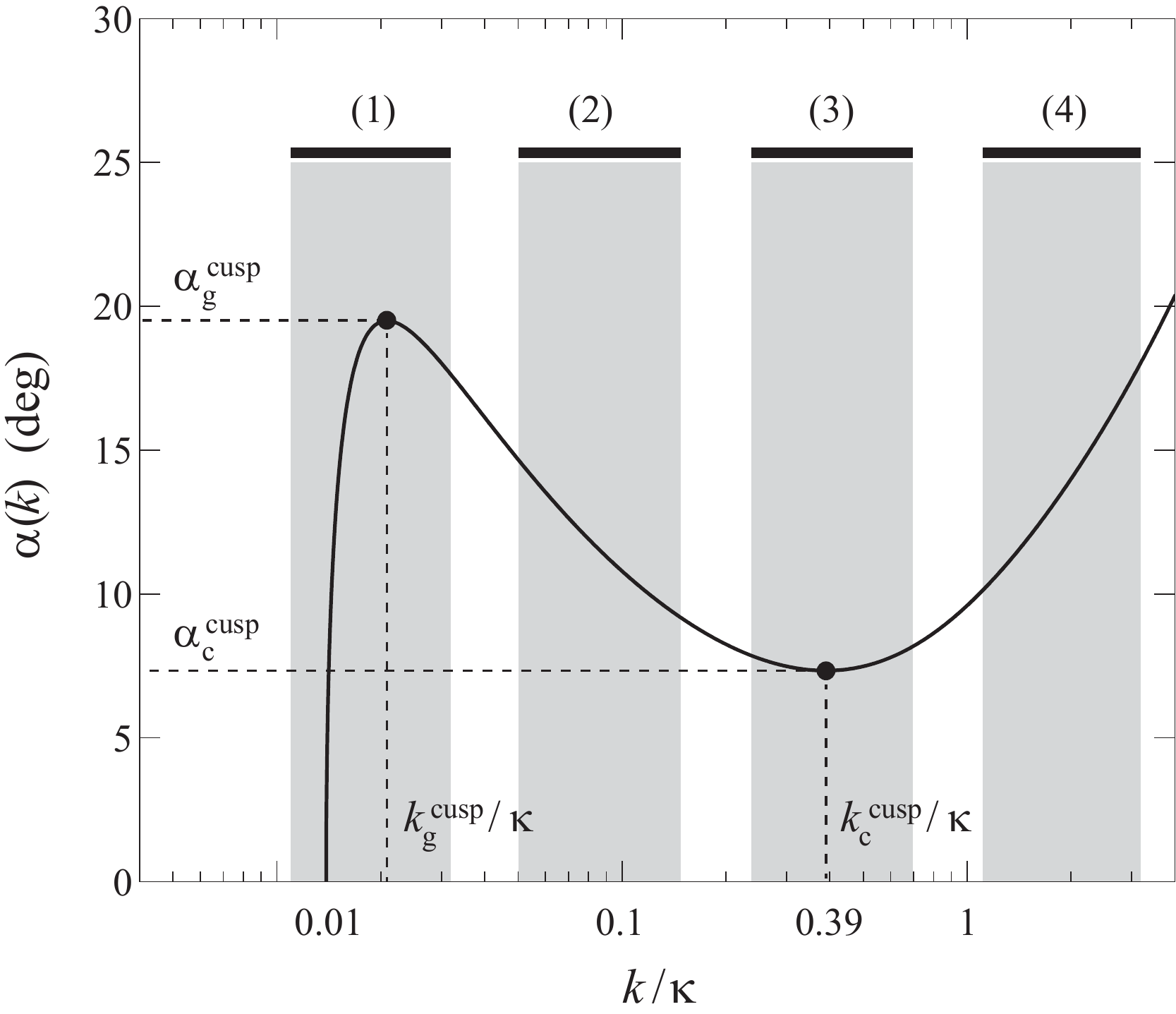}}
\caption{Radiation angle $\alpha(k)$ in the case ${\cal U} = 6$. Regions in gray represent the range of wave numbers centered on $k_f \simeq L^{-1}$ and of characteristic width $\Delta k \simeq k_f$ present in the wave packet radiated by the disturbance.  See text for the definition of the regimes $1-4$.
\label{fig:model}}
\end{figure}

\subsection{Wake regimes}

The previous analysis suggests the following picture, sketched in Fig.~\ref{fig:model}. 
Provided that ${\cal U} > {\cal U}^*$ (so the two cusp angles are defined), the following regimes may be found depending on where the energy-containing range of wave numbers $\Delta k$ centered on $k_f \simeq L^{-1}$ falls in the radiation angle curve $\alpha(k)$:

\begin{itemize}

\item[(1)] If $k_f \simeq k_g^{\rm cusp}$, the energy of the disturbance feeds the gravity cusp angle $\alpha_g^{\rm cusp}$, so the angle of maximum wave amplitude is classically given by the Kelvin angle. This regime assumes that  $\mathrm{Fr} \simeq \sqrt{k_f / k_g^{\rm cusp}}
\simeq O(1)$ and $\mathrm{Bo} \simeq {k_c^{\rm cusp}/k_f} \gg 1$.

\item[(2)] If $k_g^{\rm cusp} \ll k_f \ll k_c^{\rm cusp}$, most of the energy is radiated along the angle $\alpha(k_f)$, so the resulting wake angle is given by $\alpha = a / \mathrm{Fr}$: this is the first Mach-like regime governed by pure gravity waves. This regime holds for $\mathrm{Fr} \gg 1$, so $k_f \gg k_g^{\rm cusp}$, and $\mathrm{Bo} \gg 1$, so that $k_f \ll k_c^{\rm cusp}$.

\item[(3)] If $k_f \simeq k_c^{\rm cusp}$, the energy of the disturbance feeds the capillary cusp angle, and the resulting wake angle is given by $\alpha =  \alpha_c^{\rm cusp} \simeq c_{\rm g,min}/U = 0.77 / {\cal U}$. This is the second Mach-like regime, found for $\mathrm{Bo} \simeq O(1)$.

\item[(4)] If $k_f \gg k_c^{\rm cusp}$, the energy is radiated at arbitrary large angle, possibly in front of the disturbance ($\alpha > 90^\mathrm{o}$). However this pure capillary regime, which should be present in principle for $\mathrm{Bo} \ll 1$, is not relevant for the air-water interface because of the strong viscous attenuation at large wave numbers.

\end{itemize}

The first three regimes are compatible with the experimental wake angles reported in Fig.~\ref{fig:aexpFr}. The data at $\mathrm{Bo}_D > 2$ and $\mathrm{Fr}_D < 1.5$ correspond to regime 1, with an angle of maximum wave amplitude close to the Kelvin prediction. For $\mathrm{Fr}_D > 1.5$, the observed decrease $\alpha \simeq a/\mathrm{Fr}$, with $a \simeq 0.5$, corresponds to regime 2, similarly to that of rapid boats. Finally, the experiments at $\mathrm{Bo}_D < 2$ are compatible with regime 3, with a best fit $\alpha \simeq 0.85 /{\cal U}$ close to the prediction $0.77 / {\cal U}$. The present experimental data do not show evidence of increasing wake angle at small Bond number (regime 4), and we focus on regimes 1-3 in the following.

\subsection{Numerical simulations}
\label{sec:numsim}

\begin{figure}
\centerline{\includegraphics[width=\linewidth]{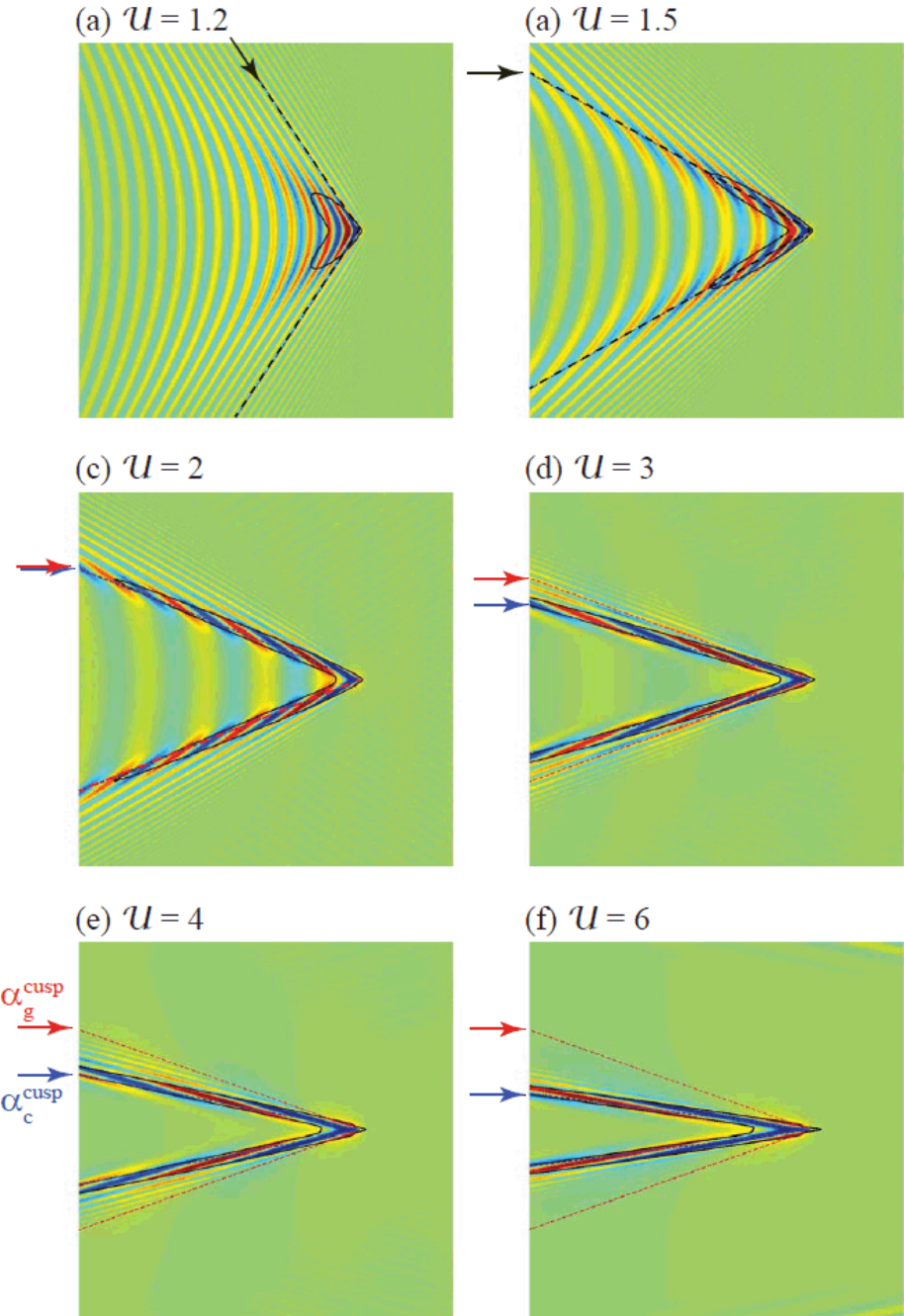}}
\caption{(Color online) Wake pattern of a Gaussian pressure disturbance at Bond number $\mathrm{Bo}=1$, for increasing velocity ${\cal U}$ between 1.2 and 6. The computation domain $L_{box}$ is $200L$, and only a subdomain of size $L_{box}/4$ is shown here. The arrows and dashed lines show the cusp precursor angle [panels (a) and (b) for ${\cal U} < {\cal U}^*$], and the gravity $\alpha_g^{\rm cusp}$ and capillary $\alpha_g^{\rm cusp}$ cusp angles [panels (c), (d), (e), and (f)]. The black contours show the isoenergy level given by 0.3 times the maximum energy.
\label{fig:simul1}}
\end{figure}

\begin{figure}
\centerline{\includegraphics[width=\linewidth]{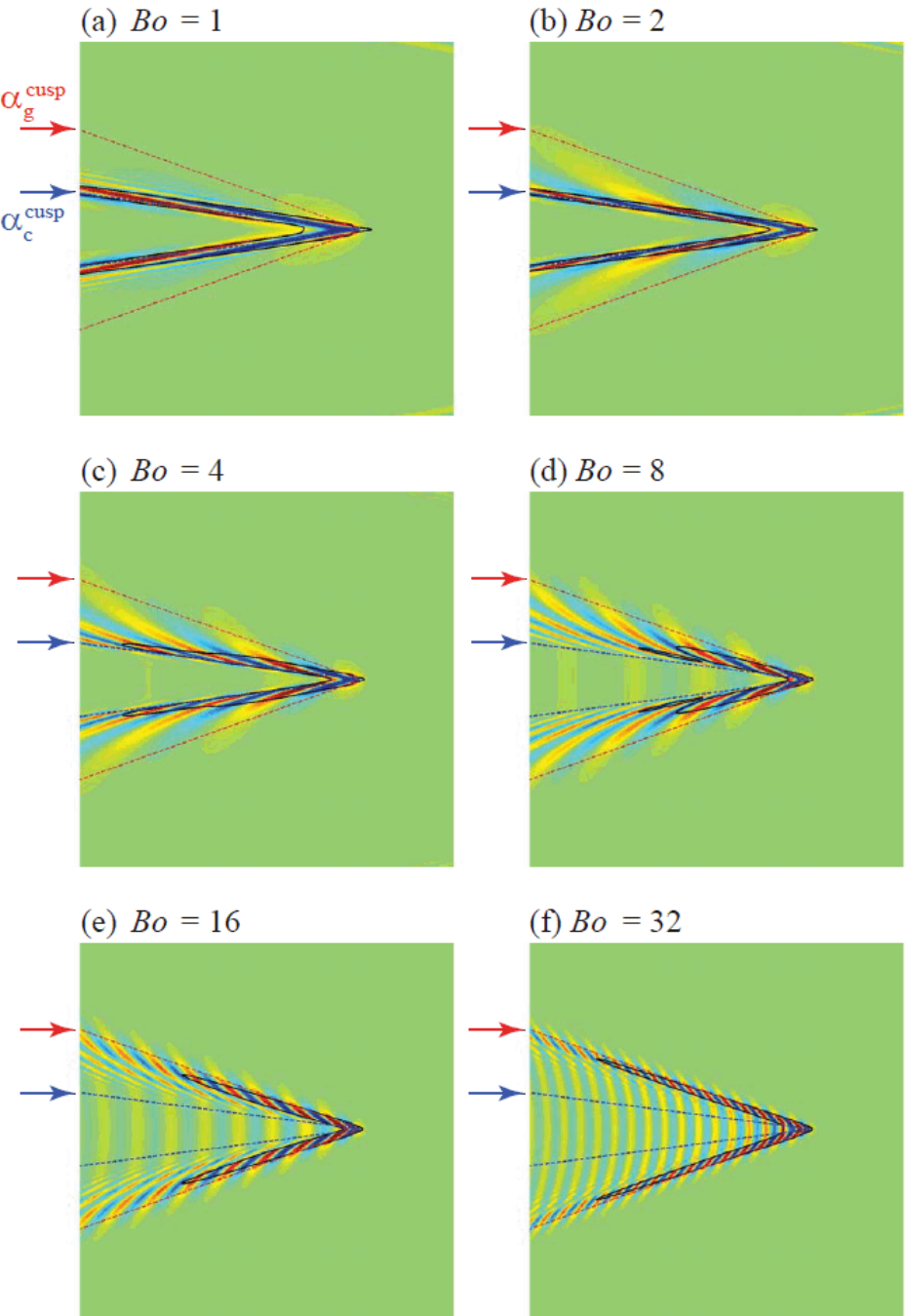}}
\caption{(Color online) Wake pattern at constant velocity  ${\cal U} = 6$, for increasing disturbance size: $\mathrm{Bo} = 1$ to 32. Same line patterns as in Fig.~\ref{fig:simul1}. As the disturbance size increases, the angle of maximum wave amplitude drifts from the capillary cusp angle $\alpha_c^{\rm cusp}$ (blue dashed line) to the gravity cusp angle $\alpha_g^{\rm cusp} \simeq \sin^{-1} (1/3)$ (red dashed line).
\label{fig:simul2}}
\end{figure}

In order to characterize the transitions between the various wake regimes, we compute the far-field angle of maximum
wave amplitude produced by an applied pressure distribution $P(r)$ traveling  at constant velocity $U$. We chose an axisymmetric Gaussian pressure distribution defined
as \cite{prefactor}
\begin{equation}
P(r) = P_0 \exp \left[ - \pi^2 \left( \frac{r}{L} \right)^2 \right],
\label{eq:p0}
\end{equation}
and we note $\mathrm{Bo} = L/\lambda_c$ and $\mathrm{Fr} = U / \sqrt{gL}$ the Bond and Froude numbers based on $L$. Assuming  linear potential flow, the surface elevation is classically obtained from the Fourier transform of the linearized Euler equation~\cite{Lighthill,Whitham},
\begin{equation}
\zeta({\bf x}) = - \lim_{\epsilon \rightarrow 0} \frac{1}{(2 \pi)^2} \int \!\!\!\! \int \frac{k \hat P ({\bf k}) / \rho}{\omega({\bf k})^2 - ({\bf k} \cdot {\bf U} - i \epsilon)^2} e^{i {\bf k} \cdot {\bf x}} d^2 {\bf k},
\label{eq:zetint}
\end{equation}
with $\hat P(k) \propto P_0 \exp [-(kL)^2/4\pi^2]$ the Fourier transform of $P(r)$. The properties of this integral have been the subject of a number of papers in the case of pure gravity waves~\cite{Wehausen}. This integral is also discussed by Lamb~\cite{Lamb} for capillary-gravity waves but is restricted to a one-dimensional wave pattern. Here we evaluate numerically Eq.~(\ref{eq:zetint}) for capillary-gravity waves on a square domain of size $L_{box}$, discretized on a grid of $N^2=8192^2$ collocation points. We set $\epsilon = 1.5 \, U / L_{box}$ as a compromise between the unphysical oscillations induced by the divergence of the integrand for small $\epsilon$ and a strong damping for large $\epsilon$. Ideally, $L_{box}$ and $N$ should be chosen such that the mesh size $L_{box}/N$ is much larger than the disturbance size and the smallest wavelength selected by the disturbance velocity $k_2$ [see Eq.~\ref{eq:k12}]. Since the range of wavenumbers satisfying the stationary condition grows rapidly as $k_2/k_1 \simeq 4 {\cal U}^4$ for large ${\cal U}$,  the full spectrum can be resolved up to ${\cal U} \simeq 4-5$ at the resolution of $N=8192$, whereas the largest wavenumbers are necessarily truncated for larger velocities. This truncation is not a limitation here, provided that the energy contained in these high wavenumbers is low, which is the case when the disturbance size is significantly larger than the mesh size.

Two series of simulations are shown in Figs.~\ref{fig:simul1} and \ref{fig:simul2} to illustrate the various wake regimes. In each panel, the dashed lines show the two cusp angles $\alpha_g^{\rm cusp}$ and $\alpha_c^{\rm cusp}$, or the cusp precursor (minimum of $\partial \alpha / \partial k$) when ${\cal U} < {\cal U}^*$, and the black line shows an iso-energy contour.

In Fig.~\ref{fig:simul1} we show the wake patterns for a small disturbance, characterized by $\mathrm{Bo} = 1$, at increasing velocity ${\cal U}$ between 1.2 and 6. Although no cusp angle is defined at ${\cal U} < {\cal U}^* = 1.938$ [Fig. 10(a) and 10(b)], a significant amount of energy concentrates in the vicinity of the cusp precursor. At ${\cal U} = 2$ [Fig. 10 (c)], slightly above the cusp onset ${\cal U}^*$, the two cusp angles are both almost equal to $21.6^\mathrm{o}$, and this is where the largest wave amplitude is found.  As ${\cal U}$ is further increased [Fig. 10(d), 10(e) and 10(f)], the cusp angles gradually separate, but since $\mathrm{Bo} = 1$ the most energetic wave number is close to the capillary wave number, so the energy concentration is mostly found around the capillary cusp (regime 3 in Fig.~\ref{fig:model}).

In Fig.~\ref{fig:simul2} we show the evolution of the wake pattern at constant velocity ${\cal U} = 6$ for increasing disturbance size ($\mathrm{Bo}$ from 1 to 32). For this particular velocity, the cusp angles are $\alpha_g^{\rm cusp} \simeq 19.50^\mathrm{o}$ (i.e., almost equal to the Kelvin angle) and $\alpha_c^{\rm cusp} \simeq 7.33^\mathrm{o}$. The wave numbers corresponding to these two cusps are well separated ($k_c^{\rm cusp} / k_g^{\rm cusp} \simeq 20$), so the three regimes can be clearly identified in this case. At small Bond number, the energy concentrates in the direction of $\alpha_c^{\rm cusp}$ [Figs.~\ref{fig:simul2}(a) and \ref{fig:simul2}(b)], in agreement with regime 3. As $\mathrm{Bo}$ is increased, the angle of maximum amplitude gradually shifts from $\alpha_c^{\rm cusp}$ to $\alpha_g^{\rm cusp}$ [Figs.~11(c), 11(d) and 11(e)], as expected for regime 2, with an energy envelope not as sharp as in Figs. 11(a) and 11(b). Finally, for the largest Bond numbers [Fig.~\ref{fig:simul2}(f)], the energy concentrates around $\alpha_g^{\rm cusp}$, and the wake pattern resembles the classical Kelvin wake (regime 1).

\begin{figure}
\centerline{\includegraphics[width=0.90\linewidth]{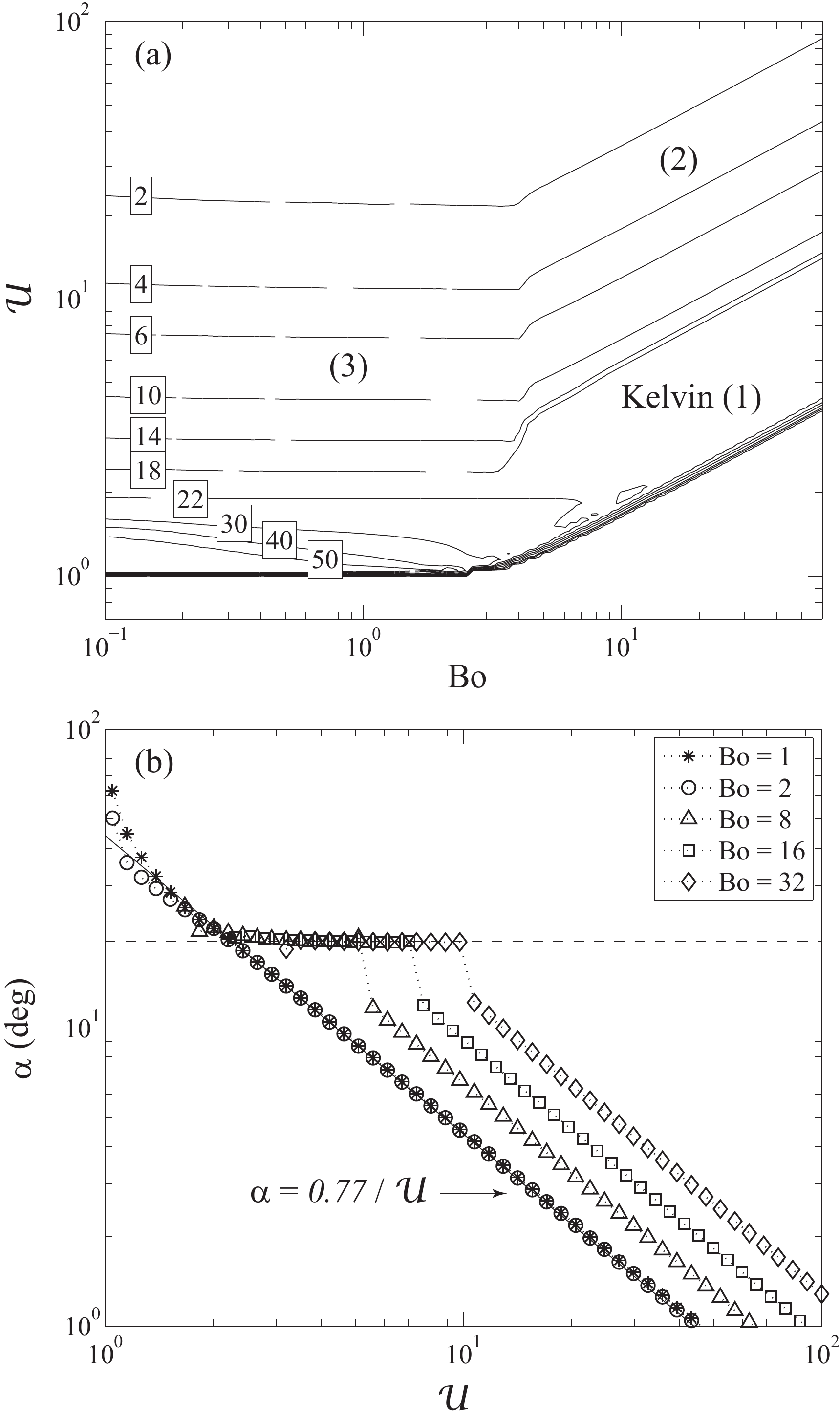}}
\caption{(a) Map of the iso-values of the angle of maximum wave amplitude $\alpha$ in the plane $(\mathrm{Bo}, {\cal U})$ for the Gaussian pressure disturbance defined by Eq.~(\ref{eq:p0}). The three wake regimes are as follows: (1) Kelvin regime, $\alpha \simeq \sin^{-1}(1/3)$; (2) Mach regime for gravity waves, $\alpha \simeq a / \mathrm{Fr}$; (3) Mach regime for capillary waves, $\alpha \simeq c_{\rm g,min} / U$. The numbers in boxes indicate the angle $\alpha$ in degrees. 
 (b) Plot of $\alpha$ as a function of ${\cal U}$, for different Bond numbers $\mathrm{Bo}$. The solid line is $\alpha = c_{\rm g,min} / U = 0.77 / {\cal U}$ (regime 3).
\label{fig:mapabou}}
\end{figure}

\begin{figure}
\centerline{\includegraphics[width=0.90\linewidth]{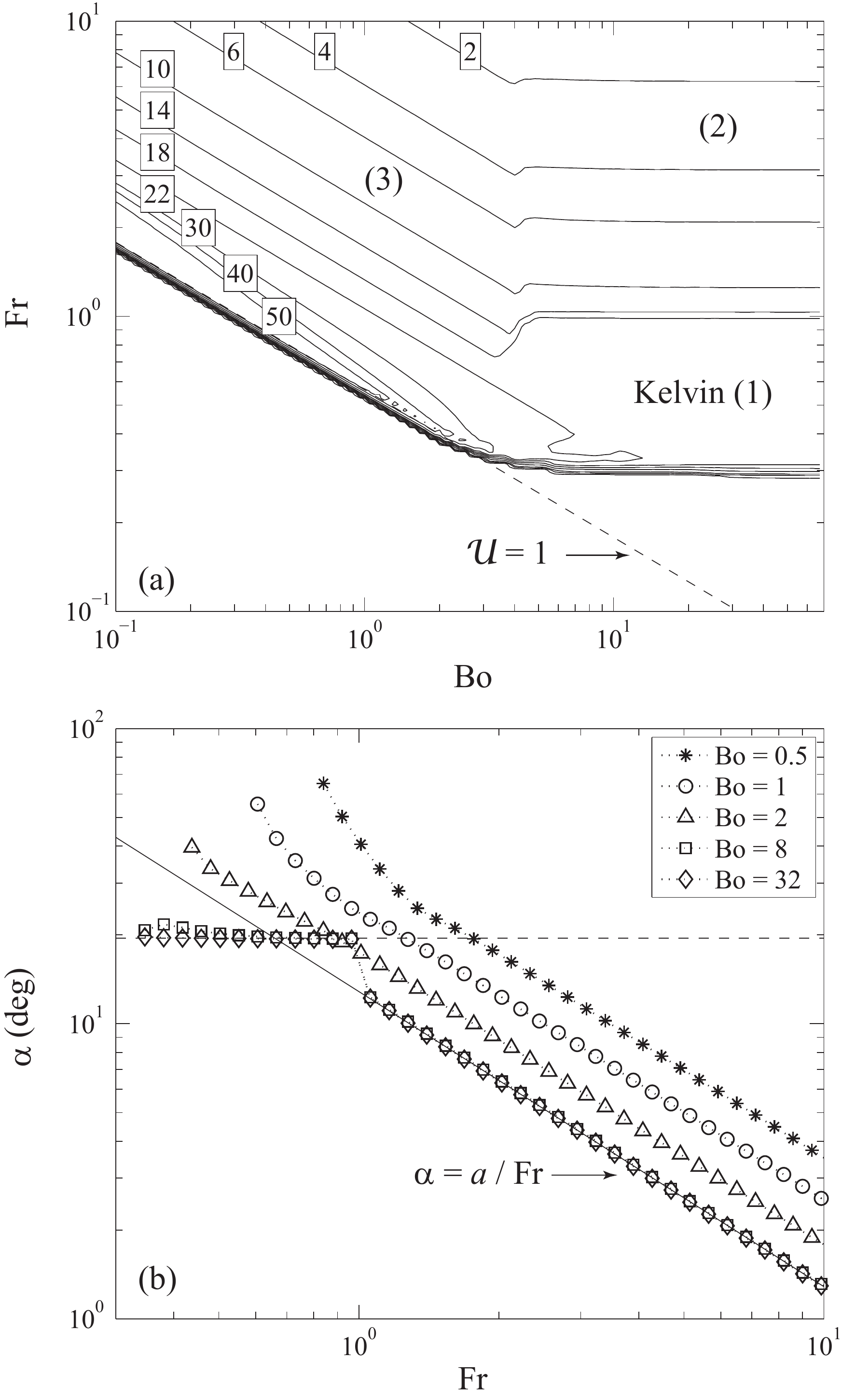}}
\caption{(a) Map of the iso-values of the angle of maximum wave amplitude $\alpha$ in the plane $(\mathrm{Bo}, \mathrm{Fr})$. The lower border ${\cal U}=1$ is given by $\mathrm{Fr} = 1/\sqrt{\pi \mathrm{Bo}}$. (b) Plot of $\alpha$ as a function of $\mathrm{Fr}$, for different values of $\mathrm{Bo}$. The solid line shows the law $\alpha = 1/(40^{1/4} \pi^{1/2} \mathrm{Fr})$ of Darmon {\it et al.}~\cite{Darmon2014} (regime 2).
\label{fig:mapabofr}}
\end{figure}

The angle of maximum wave amplitude has been systematically measured for $\mathrm{Bo} \in [0.1, 60]$ and ${\cal U} \in [1, 100]$, and the results are summarized in Figs.~\ref{fig:mapabou}  and \ref{fig:mapabofr}. Using the set of parameters $(\mathrm{Bo}, {\cal U})$, the iso-$\alpha$ curves are independent of $\mathrm{Bo}$ for $\mathrm{Bo}<3$ [Fig.~\ref{fig:mapabou}(a)], and the decrease of $\alpha$ with velocity is in excellent agreement with the law $c_{\rm g,min}/U$ at large velocity [Fig.~\ref{fig:mapabou}(b)]. At small velocity, $\alpha$ approximately follows the precursor cusp angle shown in Fig.~\ref{fig:cusp}(a). Plotting now the same data in terms of the parameters $(\mathrm{Bo}, \mathrm{Fr})$, we find that the iso-$\alpha$ curves are now independent of $\mathrm{Bo}$ for $\mathrm{Bo}> 4$ [Fig.~\ref{fig:mapabofr}(a)], and the law $\alpha = a/\mathrm{Fr}$ with $a=1 / (40^{1/4} \pi^{1/2} \mathrm{Fr})$ of Darmon {\it et al}.~\cite{Darmon2014} for pure gravity waves is accurately recovered [Fig.~\ref{fig:mapabofr}(b)]. Considering that the transition Bond number corresponds to a disturbance size $L$ equal to the wavelength of minimum group velocity $2\pi / k_{c}^{\rm cusp} \simeq 2.54 \lambda_c$ simply predicts $Bo_c \simeq 2.54$, which is close to the actual transition.

We can conclude that the overall behavior of the wake angles computed numerically confirms the picture
given in Fig.~\ref{fig:model}. Of course, the agreement with the experimental measurements of Fig.~\ref{fig:aexpFr} remains qualitative: the complex flow around a bluff body cannot be reduced to a simple pressure disturbance characterized by a single scale. In particular, the transition Bond number between regimes 1-2 and regime 3 for the pressure disturbance ($Bo_c \simeq 4$) is significantly larger than the experimental one ($Bo_{D,c} \simeq 0.7$). This discrepancy is consistent with the argument proposed in Sec.~\ref{sec:ewa}: The ratio between the experimental and the numerical transition Bond number suggests that a cylinder of diameter $D$ has an effect comparable to a Gaussian pressure distribution of size $L \simeq 6 D$.  Another difference is the presence of sharp jumps of $\alpha$ in the simulations, when the maximum wave amplitude switches from the gravity cusp (Kelvin angle) to the intermediate regime 2 ($\alpha \simeq 1/\mathrm{Fr}$), whereas a smooth transition is found in the experimental data.   In spite of these differences, it is remarkable that the scaling laws for the three wake regimes identified experimentally could be well reproduced by the present model and simulations.

\section{Conclusion}

We have shown that the decrease with velocity of the angle of maximum wave amplitude, found in Ref.~\cite{Rabaud2013} for ship wakes in the gravity regime, is also present for the capillary-gravity wakes generated by a disturbance of size comparable to the capillary length. In all cases, the wake angle is found to decrease following a law in the form $c_g/U$ at large velocity, as in the Mach cone problem, where the ``sound velocity'' $c_g$ is the group velocity of the dominant wavepacket excited by the disturbance. At large Bond number (weak capillary effects), $c_g$ corresponds to the group velocity of the gravity waves of wavelength comparable to the disturbance size, whereas at Bond number of order unity (large capillary effects) it is given by the minimum group velocity of capillary-gravity waves, $c_{\rm g,min} \simeq 0.77 c_{\rm min}$. Using the general property of dispersive waves that the waves of maximum amplitude excited by a disturbance have their wavelength of order of the disturbance size, we provide a simple linear model based on an applied pressure disturbance which describes the transition between the Kelvin regime and the two Mach-like regimes.  Although the complex flow phenomena present in the experiments (detached boundary layers, vortex shedding, wave breaking, turbulence) cannot be accounted for by such a pressure disturbance, it is remarkable that this simple model reproduces with reasonable accuracy the behavior of the far-field wake angle. Note that although the angle of maximum wave amplitude follows a Mach-like law, the problem remains dispersive in nature, which is illustrated by the fact that the crest angle (governed by the phase velocity) never coincides with the wake angle (governed by the group velocity).

Amusingly, we note that the wake behind a duck, often used to illustrate the universal properties of the Kelvin wake pattern, nearly falls in the complex intermediate situation where $\mathrm{Bo} \simeq O(1)$, ${\cal U} \simeq O(1)$ and $\mathrm{Fr} \simeq O(1)$.

\appendix

\section{Influence of the wake angle definition}

The wake angle in the model and in the numerical simulations of Sec.~\ref{sec:fse} is defined as the angle of maximum wave amplitude. On the other hand, the visualization methods used in the experiments of Sec.~\ref{sec:setup} and in the analysis of the airborne images of ship wakes in Ref.~\cite{Rabaud2013} are not based on the wave amplitude, but rather on the wave slope or curvature, which may introduce a bias. It is therefore important to check the robustness of the results with respect to the definition used for the wake angle.

We have simulated the wake pattern of a Gaussian pressure disturbance in the pure gravity regime (large $\mathrm{Bo}$) following the method described in Sec.~\ref{sec:numsim}, and determined the wake angle according to the following definitions:

1. Maximum of wave amplitude $\zeta$. This is the reference definition, which is used in Sec.~\ref{sec:numsim} and in Refs.~\cite{Rabaud2013,Darmon2014,Ellingsen2014,Moisy2014,Benzaquen2014,Pethi2014}.

2. Maximum of longitudinal wave slope $\partial \zeta / \partial x$. This definition is relevant to the swimming-pool experiments, in which the wake angle is determined from reflection of natural light (Fig.~\ref{fig:pool}).

3. Maximum of lateral wave slope $\partial \zeta / \partial y$.

4. Maximum of absolute wave slope $|\nabla \zeta|$.

5. Maximum of curvature $\nabla^2 \zeta$. This definition is relevant to the shadowgraphy visualization used in the small-scale experiments (Fig.~\ref{fig:fast}).

\begin{figure}
\centerline{\includegraphics[width=0.90\linewidth]{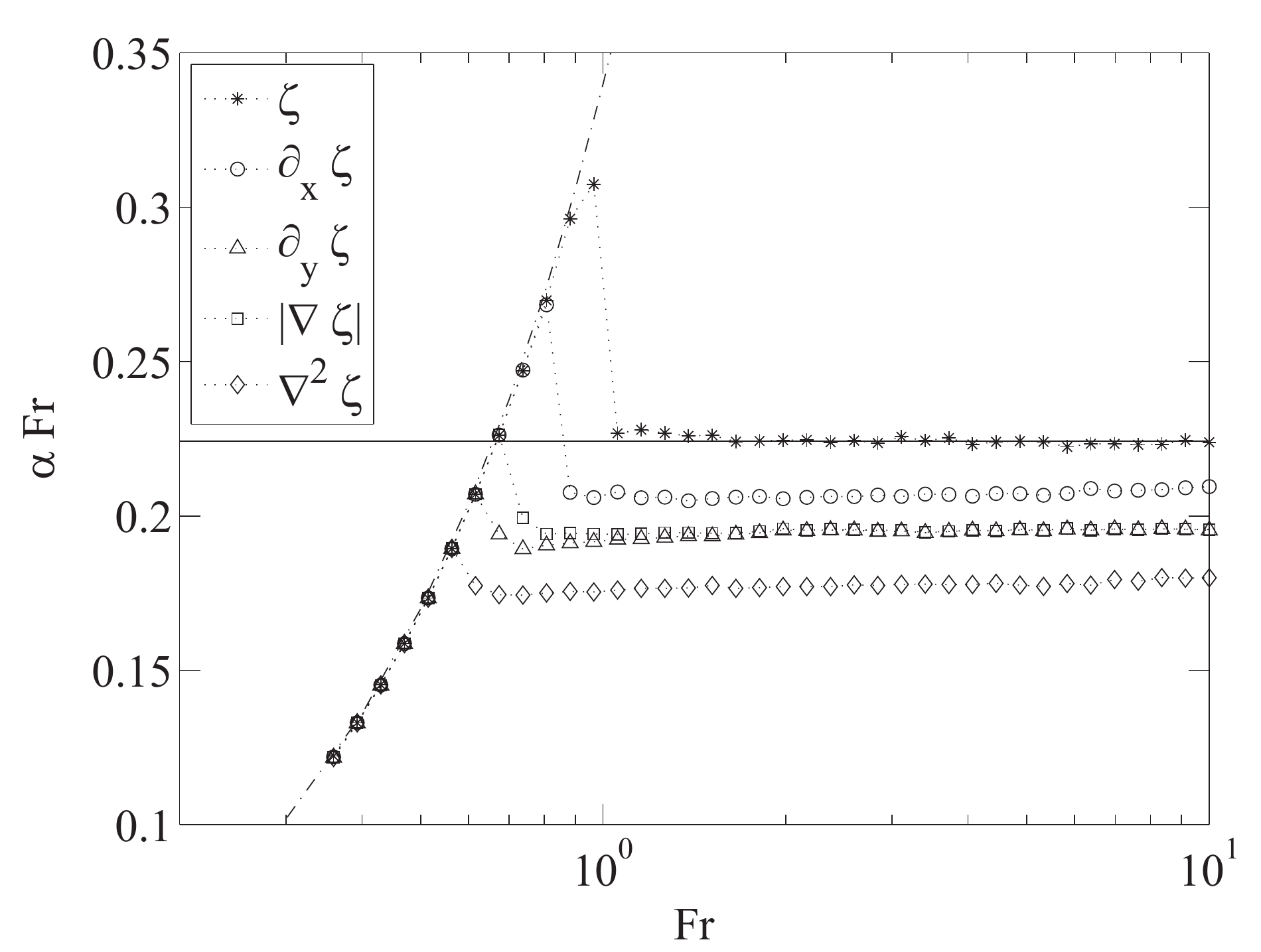}}
\caption{Compensated angle $\alpha \mathrm{Fr}$ as a function of $\mathrm{Fr}$ for a Gaussian pressure disturbance using various definitions for the wake angle. The dash-dotted curve shows the Kelvin regime $\alpha_K \mathrm{Fr}$, with $\alpha_K = \sin^{-1}(1/3)$. The reference definition ($\ast$), corresponding to the angle of maximum wave amplitude $\zeta$, is compared to the exact value $a = 1/(40^{1/4} \pi^{1/2}) \simeq 0.224$ (horizontal line). The four alternate definitions, based on maximum wave slope ($\circ$, $\triangle$, and $\square$) and curvature ($\diamond$), give smaller values of $a$. 
\label{fig:bias}}
\end{figure}

For each definition, the measured wake angle is equal to the Kelvin angle $\alpha_K = \sin^{-1}(1/3)$ at small Froude number and decreases as $\alpha \simeq a/\mathrm{Fr}$ at larger $\mathrm{Fr}$, indicating that this transition is not sensitive to the exact definition of the wake angle, at least in the case of a Gaussian pressure disturbance. However, a dependence of the prefactor $a$ is found depending on the  definition used, as shown when plotting the compensated angle $\alpha \mathrm{Fr}$ (Fig.~\ref{fig:bias}). For the reference definition 1 we recover the exact value $a_1 \simeq 1/(40^{1/4} \pi^{1/2}) \simeq 0.224$ of Ref.~\cite{Darmon2014}, but smaller values are found for the other definitions as follows: $a_2 \simeq 0.207 (-8\%)$, $a_3 \simeq a_4 \simeq 0.195 (-13\%)$ and $a_5 \simeq 0.178 (-21\%)$ (definitions 3 and 4 give essentially the same result because at large $\mathrm{Fr}$ the dominant contribution to the wave slope is in the transverse direction $y$). If we consider, for instance, $\mathrm{Fr} = 2$ (corresponding to a cylinder-based Froude number $\mathrm{Fr}_D \simeq 5$), the angle of maximum wave amplitude is $6.4^\mathrm{o}$, the angle of maximum slope is $5.6^\mathrm{o}$, and the angle of maximum curvature is $5.1^\mathrm{o}$. These differences are comparable to the experimental uncertainties in Fig.~\ref{fig:aexpFr}, suggesting that the present results are not significantly affected by these measurement biases.

\acknowledgments
We acknowledge M. Chestier and the staff of the swimming pool of Orsay, and S. Atis, A. Aubertin, L. Auffray, C. Borget, P.-P. Cortet, R. Pidoux and B. Saintyves for experimental help. We thank G. Rousseaux for pointing out key references. F.M. acknowledges Institut Universitaire de France.

\end{document}